%% file: main.tex
\documentclass[10pt,twocolumn,letterpaper]{article}

\usepackage[pagenumbers]{cvpr}            

\input{preamble}

\definecolor{cvprblue}{rgb}{0.21,0.49,0.74}
\usepackage{xr}       
\usepackage{booktabs} 
\usepackage{multirow} 
\usepackage{graphicx} 
\usepackage{makecell}
\usepackage[accsupp]{axessibility}
\usepackage[pagebackref,breaklinks,colorlinks,allcolors=cvprblue]{hyperref}
\usepackage{tikz}
\usetikzlibrary{positioning}

\usepackage{booktabs} 
\usepackage{multirow} 
\usepackage{graphicx} 
\def\paperID{14283} 
\def\confName{CVPR}
\def\confYear{2026}

\title{cryoSENSE: Compressive Sensing Enables High-throughput Microscopy with Sparse and Generative Priors on the Protein Cryo-EM Image Manifold}

\author{Zain Shabeeb\thanks{Equal contribution.}\quad
Daniel Saeedi\footnotemark[1]\quad
Darin Tsui\footnotemark[1]\quad
Vida Jamali\thanks{Co-corresponding authors.}\quad
Amirali Aghazadeh\footnotemark[2]\\
Georgia Institute of Technology\\
225 North Ave NW, Atlanta, Georgia 30332\\
{\tt\small \{zshabeeb3, dsaeedi3, darint, vida, amiralia\}@gatech.edu}
}

\begin{document}
\maketitle
\input{sec/0_abstract}    
\input{sec/1_intro}

\input{sec/2_background}

\input{sec/3_results}

\section*{Acknowledgments}
This research was supported by the Exponential Electronics seed grant of the Institute for Matter and Systems at Georgia Tech.  A.A. and D.S. acknowledge the support of Microsoft made available via GT Cloud Hub. D.T. acknowledges the support of the National Science Foundation (NSF) Graduate Research Fellowship Program (GRFP). The authors acknowledge the support of the Material Characterization Facility and the Electron Microscopy Facility of the Institute for Matter and Systems at Georgia Tech, a member of the National Nanotechnology Coordinated Infrastructure (NNCI), which is supported by the National Science Foundation (ECCS-2025462).

{
    \small
    \bibliographystyle{ieeenat_fullname}
    \bibliography{main}
}
\newpage
\clearpage
\input{Appendix/supp}


\end{document}

%% file: preamble.tex
\newcommand{\TODO}[1]{\textbf{\color{red}[TODO: #1]}}
\usepackage{algorithm}
\usepackage{algpseudocode}
\usepackage{float}
\DeclareMathOperator*{\argmin}{arg\,min}
\renewcommand{\TODO}[1]{}

\usepackage{microtype}



%% file: sec/0_abstract.tex
\begin{abstract}

Cryo-electron microscopy (cryo-EM) enables the atomic-resolution visualization of biomolecules; however, modern direct detectors generate data volumes that far exceed the available storage and transfer bandwidth, thereby constraining practical throughput. We introduce cryoSENSE, the computational realization of a hardware-software co-designed framework for compressive cryo-EM sensing and acquisition. We show that cryo-EM images of proteins lie on low-dimensional manifolds that can be independently represented using sparse priors in predefined bases and generative priors captured by a denoising diffusion model. cryoSENSE leverages these low-dimensional manifolds to enable faithful image reconstruction from spatial and Fourier-domain undersampled measurements while preserving downstream structural resolution. In experiments, cryoSENSE increases acquisition throughput by up to 2.5$\times$ while retaining the original 3D resolution, offering controllable trade-offs between the number of masked measurements and the level of downsampling. Sparse priors favor faithful reconstruction from Fourier-domain measurements and moderate compression, whereas generative diffusion priors achieve accurate recovery from pixel-domain measurements and more severe undersampling.\footnote{Project website: \href{https://cryosense.github.io}{https://cryosense.github.io}}

\end{abstract}

%% file: sec/1_intro.tex
\section{Introduction}
\label{sec:intro}

Single particle cryogenic electron microscopy (cryo-EM) is a crucial tool for structural biology~\cite{yip2020atomic, nakane2020single}, enabling the determination of three-dimensional (3D) structures of biomolecular complexes at near-atomic resolution and thereby helping to elucidate the complex structure-function relationships of proteins~\cite{nogales2016development}. In a cryo-EM experiment, tens to hundreds of thousands of two-dimensional (2D) images are acquired from biomolecules within a flash-frozen specimen using a direct electron detector on a transmission electron microscope. High-throughput cryo-EM detectors generate raw data streams of several gigabytes per second~\cite{datta2021data, chua2022better, poger2023big}, quickly overwhelming local storage and network bandwidth. In current workflows, three strategies are mainly used to manage this load. The first approach, common in single-particle cryo-EM, involves acquiring many short subframes for electron counting, then summing them before saving to disk, which reduces the data rate but sacrifices temporal resolution~\cite{guo2020electron}. The second is to shorten total acquisition time, filling local storage within minutes and then idling the microscope while data are offloaded to network drives~\cite{meng2023best, poger2023big}. The third relies on post-acquisition compression~\cite{eng2019reducing,datta2021data}, which reduces storage requirements retrospectively but does not mitigate real-time bandwidth or throughput bottlenecks.

\begin{figure}
    \centering
    \includegraphics[width=0.9\linewidth]{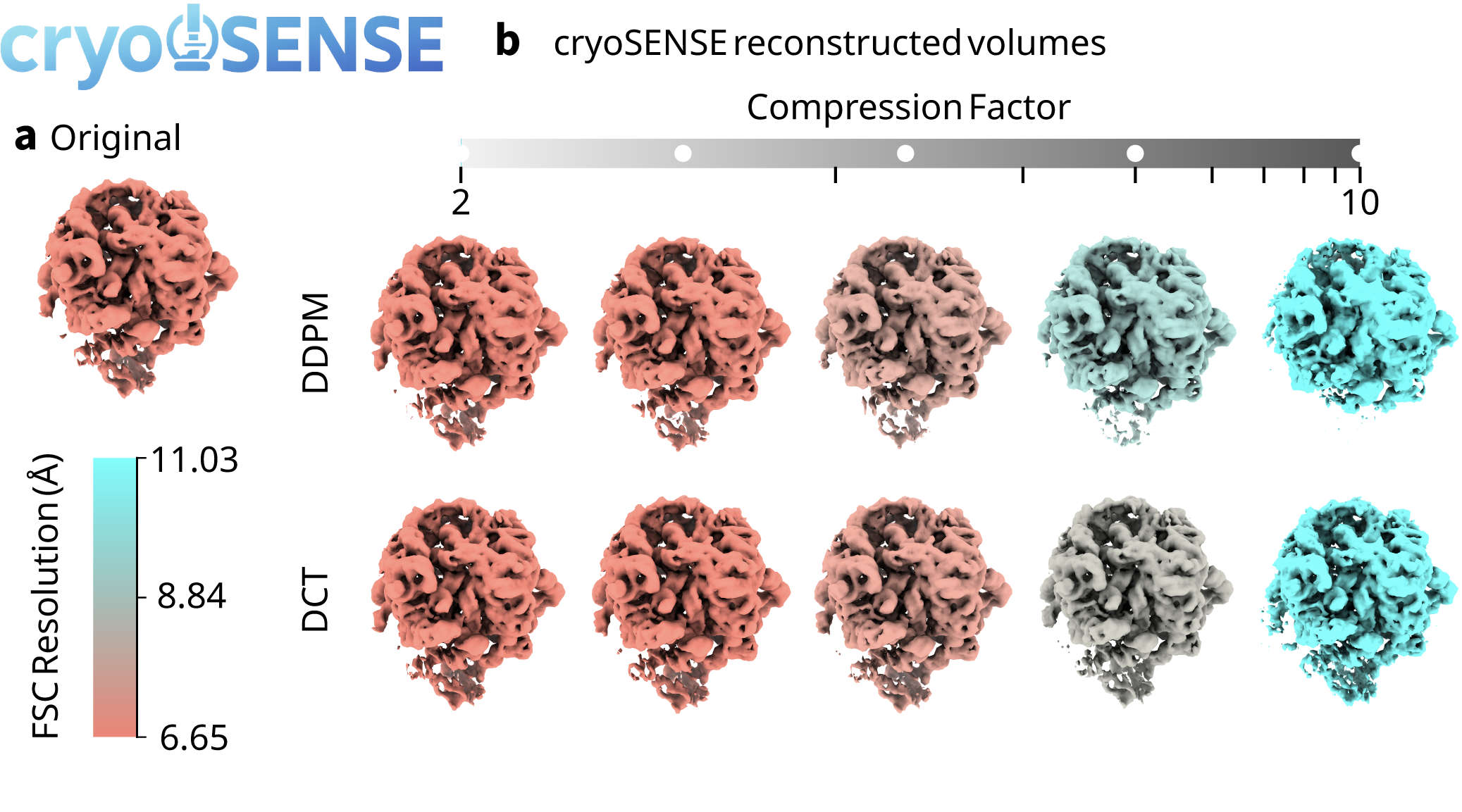}
    \caption{\textbf{cryoSENSE increases data throughput by preserving 3D structural detail under high compression factors.} \textbf{a}, Original 3D cryo-EM volume from uncompressed particle images. \textbf{b}, cryoSENSE 3D reconstructions from compressively acquired images, using generative (top row) and sparse priors (bottom row). Volumes are color-coded by their FSC resolution (lower, better).}
    \vspace{-0.5cm}
    \label{fig:placeholder}
\end{figure}

\begin{figure*}[t!]
    \centering
    \includegraphics[width=0.75\linewidth]{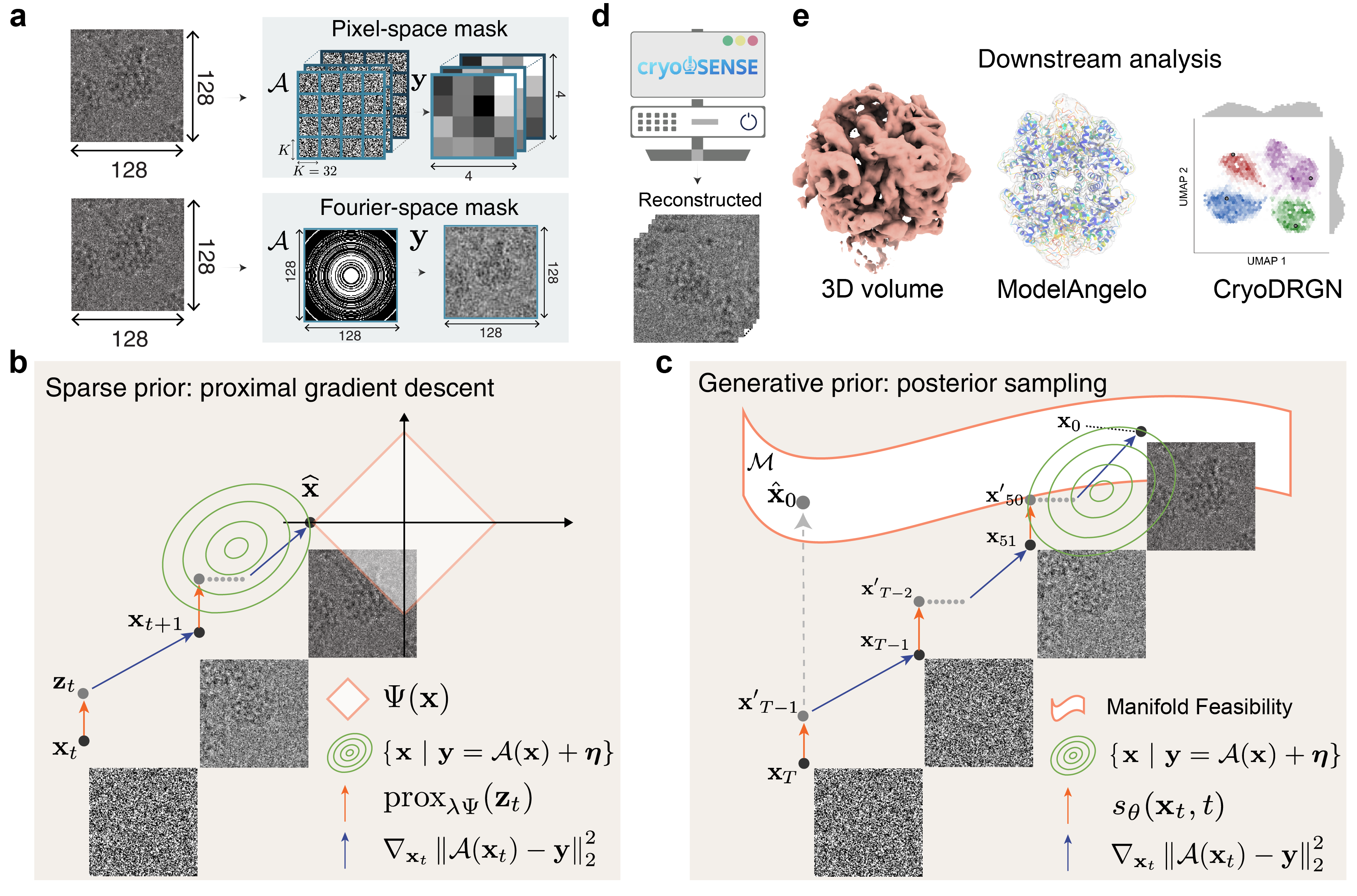}
    \caption{\textbf{Overview of cryoSENSE.} \textbf{a}, cryoSENSE employs pixel-space and Fourier-space masking strategies to obtain compressed measurements. \textbf{b}, Sparsity priors enable image recovery via proximal gradient descent. \textbf{c}, Generative priors learn a low-dimensional manifold of cryo-EM images and guide a diffusion process to generate images consistent with the measurements. \textbf{d}, cryoSENSE enables high-throughput acquisition of cryo-EM images, which are then validated through downstream biological tasks such as \textbf{e}, 3D volume reconstruction, atomic model building, and conformational heterogeneity analysis.}
    \label{fig:1}
\end{figure*}

Fortunately, raw cryo-EM image data are highly structured, suggesting that a compressive acquisition paradigm could fully leverage these structures to capitalize on the capabilities of modern detectors. Herein, we introduce cryoSENSE, a computational demonstration of a hardware–software co-designed framework for compressive cryo-EM sensing and acquisition. cryoSENSE collects compressed measurements during imaging, substantially increasing data throughput, and computationally reconstructs high-fidelity 2D images that preserve downstream structural 3D resolution (Fig.~\ref{fig:placeholder}).

Recovering images from compressed measurements is a fundamentally ill-posed inverse problem, as infinitely many candidate images can be consistent with the same compressed observations. To address this challenge, cryoSENSE employs two complementary strategies~\cite{donoho2006compressed, candes2006robust, candes2006stable}.
First, cryoSENSE imposes handcrafted priors that constrain the solution space~\cite{yu2011solving}, particularly sparsity in predefined transform domains (e.g., wavelets or discrete cosine transform bases)~\cite{rao2014discrete}, or regularization schemes such as total variation, which promote piecewise smoothness~\cite{vogel1996iterative}. Such priors have demonstrated effectiveness across diverse imaging modalities, including magnetic resonance imaging~\cite{lustig2007sparse}, infrared and visible-wavelength imaging~\cite{asif2016flatcam}, four-dimensional scanning transmission electron microscopy~\cite{nicholls2022compressive, ni2024framework, robinson2023simultaneous}, and fluorescence microscopy~\cite{ma2021high, calisesi2022compressed}.
Second, cryoSENSE learns data-driven priors directly from experimental images using generative models~\cite{choi2021ilvr, wang2024dmplug, chung2023diffusion, ravula2023optimizing, song2022solving}. In particular, it leverages denoising diffusion probabilistic models (DDPMs)~\cite{ho2020denoisingdiffusionprobabilisticmodels, dhariwal2021diffusion}, replacing the assumption of sparsity in a fixed basis with the assumption that images lie on a low-dimensional manifold~\cite{bora2017compressed, jalal2020robust, asim2020blind, evans2025cryo, lee20253d}. By learning to reverse a stochastic noising process trained on real cryo-EM data, diffusion models are guided to generate high-quality reconstructions from incomplete or corrupted measurements~\cite{choi2021ilvr, wang2024dmplug, chung2023diffusion, song2024solving, lugmayr2022repaint}.

We demonstrate the modularity of cryoSENSE using two broad categories of sampling schemes: pixel-space masking, which can be realized using a physical or nanofabricated pattern that modulates transmission before electrons reach the detector (i.e., a coded aperture), and Fourier-space masking, which can be realized using modulation in the back focal plane (phase plates, holographic gratings) that mixes spatial frequencies before detection.

\noindent {\bf Timeliness.} The exponential explosion of the Electron Microscopy Public Image Archive (EMPIAR)~\cite{iudin2023empiar} provides unprecedented data for learning stronger generative priors. As of today, EMPIAR contains more than 2,508 entries and occupies approximately 6.9 PiB of storage, with new data added daily. Moreover, recent advances in generative diffusion models have transformed inverse problem solving in imaging~\cite{song2022solving, daras2024survey, hu2024learning, chung2025diffusion}, enabling the generation of high-quality images. In parallel, advances in programmable electron-optical elements have made Fourier-domain modulation of the electron beam experimentally viable. Electrostatic and MEMS-based phase plates~\cite{grillotoward, verbeeck2018demonstration, rosi2022theoretical} now enable user-defined phase and amplitude modulation of the beam directly in the back focal plane, effectively realizing Fourier-space masking prior to sensing. Modern cryo-EM direct electron detectors, including Gatan K3, Thermo Fisher Falcon 4i, and direct electron DE-64, further enable binning (i.e., combining adjacent sensor pixels or electron events into larger effective pixels) as a user-configurable mode, allowing spatial resolution to be traded for higher data throughput (and lower storage demand). 

\noindent\textbf{Contributions.}
We introduce cryoSENSE, the first computational demonstration of a compressive cryo-EM protein imaging. Our main contributions are:
\begin{itemize}[leftmargin=10pt]

\item \textbf{A new framework for compressive cryo-EM acquisition.}
cryoSENSE supports both pixel-space and Fourier-space sampling and reconstructs protein images using either handcrafted sparsity priors or learned generative diffusion priors.

\item \textbf{Quantitative limits of compressed cryo-EM protein imaging.} We establish the first empirical operating regime for compressed cryo-EM, showing 2D particle images can be faithfully reconstructed at compression factors up to 2.7$\times$ in pixel-space and 2.5$\times$ in Fourier-space.

\item \textbf{Sparsity and generative tradeoffs.} We present an empirical tradeoff for compressed cryo-EM imaging: sparse priors favor faithful reconstruction from Fourier-space sampling and moderate compression factors, whereas generative priors favor faithful reconstruction from pixel-space sampling and more extreme downsampling at higher compression factors.

\item \textbf{High-fidelity 3D volume reconstruction.} We demonstrate that cryoSENSE exhibits faithful 3D volume reconstruction at compression factors up to $1.5\times$ in pixel-space and $2.5\times$ in Fourier-space, recovering near-perfect Fourier Shell Correlation (FSC) resolution.

\item \textbf{Preservation of biologically meaningful protein structure.} cryoSENSE reconstructions retain downstream biological signal, preserving conformational heterogeneity with 80–88\% cluster agreement in CryoDRGN and atomic model building with backbone RMSDs of 2.1–2.3~\AA{} in ModelAngelo.

\end{itemize}

%% file: sec/2_background.tex
\section{Method}

\begin{table*}[t]
    \centering
    \caption{
    Average LPIPS and SSIM scores for cryoSENSE pixel-space masking reconstructions across downsampling level $K$ and compression factor $C$ (rounded to the nearest tenth). 
    }
    \label{tab:real_space}
    \resizebox{\textwidth}{!}{%
    \begin{tabular}{l|c|cccc|cccc|cccc|cccc|cccc}
    \toprule
    \multirow{2}{*}{Prior} & $K$ & \multicolumn{4}{c|}{$2$} 
    & \multicolumn{4}{c|}{$4$} 
    & \multicolumn{4}{c|}{$8$} 
    & \multicolumn{4}{c|}{$16$} 
    & \multicolumn{4}{c}{$32$} \\
    \cmidrule(lr){2-22}
     & $C$ & 4 & 2 & 1.3 & 1 
     & 16 & 2.7 & 1.5 & 1 
     & 10.7 & 2.6 & 1.5 & 1 
     & 10.2 & 2.5 & 1.4 & 1 
     & 10 & 2.5 & 1.4 & 1 \\
    \midrule
    \multirow{2}{*}{Sparse - DCT}
     & LPIPS ($\downarrow$) & 0.16 & \textbf{0.11} & \textbf{0.08} & \textbf{0.05} & 0.27 & 0.14 & 0.09 & 0.05 & 0.26 & 0.16 & 0.13 & 0.10 & 0.30 & 0.25 & 0.28 & 0.25 & 0.40 & 0.40 & 0.39 & 0.36 \\
     & SSIM ($\uparrow$) & \textbf{0.38} & 0.59 & 0.75 & 0.86 & 0.14 & 0.50 & 0.75 & 0.91 & 0.16 & 0.49 & \textbf{0.68} & 0.79 & \textbf{0.15} & \textbf{0.36} & 0.42 & 0.42 & 0.05 & 0.06 & 0.14 & 0.18 \\
    \midrule
    \multirow{2}{*}{Sparse - WT}
     & LPIPS ($\downarrow$) & 0.21 & 0.13 & 0.09 & 0.05 & 0.44 & 0.16 & 0.09 & 0.05 & 0.39 & 0.22 & 0.14 & 0.10 & 0.45 & 0.33 & 0.33 & 0.26 & 0.45 & 0.42 & 0.39 & 0.36 \\
     & SSIM ($\uparrow$) & 0.38 & 0.59 & 0.74 & 0.86 & \textbf{0.18} & \textbf{0.49} & \textbf{0.75} & 0.90 & \textbf{0.16} & \textbf{0.47} & 0.67 & 0.78 & 0.02 & 0.34 & 0.40 & 0.42 & 0.02 & 0.09 & 0.14 & 0.18 \\
    \midrule
    \multirow{2}{*}{Sparse - TV}
     & LPIPS ($\downarrow$) & 0.37 & 0.20 & 0.10 & 0.06 & 0.38 & 0.23 & 0.14 & \textbf{0.04} & 0.38 & 0.32 & 0.19 & \textbf{0.07} & 0.36 & 0.31 & 0.22 & 0.14 & 0.36 & 0.32 & 0.27 & 0.26 \\
     & SSIM ($\uparrow$) & 0.03 & \textbf{0.64} & \textbf{0.78} & \textbf{0.88} & 0.02 & 0.52 & 0.73 & \textbf{0.92} & 0.02 & 0.20 & 0.53 & \textbf{0.85} & 0.04 & 0.20 & 0.43 & 0.63 & 0.04 & 0.15 & 0.25 & 0.28 \\
    \midrule
    \multirow{2}{*}{Gen - DDPM}
     & LPIPS ($\downarrow$) & \textbf{0.14} & 0.12 & 0.09 & 0.06 & \textbf{0.27} & \textbf{0.12} & \textbf{0.09} & 0.06 & \textbf{0.20} & \textbf{0.15} & \textbf{0.11} & 0.07 & \textbf{0.21} & \textbf{0.17} & \textbf{0.12} & \textbf{0.10} & \textbf{0.23} & \textbf{0.18} & \textbf{0.17} & \textbf{0.16} \\
     & SSIM ($\uparrow$) & 0.31 & 0.50 & 0.68 & 0.81 & 0.18 & 0.44 & 0.68 & 0.84 & 0.15 & 0.15 & 0.43 & 0.66 & 0.15 & 0.44 & \textbf{0.62} & \textbf{0.72} & \textbf{0.17} & \textbf{0.41} & \textbf{0.48} & \textbf{0.53} \\
    \bottomrule
    \end{tabular}
    }
\end{table*}

\subsection{cryoSENSE: Inverse Problem Formulation} In cryoSENSE, we solve the inverse problem of recovering a cryo-EM image $\mathbf{x}^* \in \mathbb{R}^n$ from noisy, compressed measurements $\mathbf{y} = \mathcal{A}(\mathbf{x}^*) + \boldsymbol{\eta} \in \mathbb{R}^m$, where $\mathcal{A} : \mathbb{R}^n \to \mathbb{R}^m$ is a known linear projection operator (in pixel or Fourier space) representing the acquisition of $m \ll n$ masked images and $\boldsymbol{\eta} \sim \mathcal{N}(\mathbf{0}, \sigma^2 \mathbf{I})$ represents additive white Gaussian noise (Fig.~\ref{fig:1}a). To solve this ill-posed problem, we assume two different priors over cryo-EM images: i) sparsity in a pre-defined basis and ii) living on a low-dimensional manifold $\mathcal{M}$ learned from the EMPIAR data. In the former case, we use proximal gradient descent, and in the latter, we do posterior sampling on a denoising diffusion model to recover the protein image $\mathbf{x}^*$.

\subsection{Image Recovery with Sparse Priors}

To impose the sparsity prior in cryoSENSE, we formulate the following convex optimization problem:  
\begin{equation}
\mathbf{\widehat{x}} = \argmin_{\mathbf{x}} \quad \left\| \mathcal{A}(\mathbf{x}^*) - \mathbf{y} \right\|_2^2 + \lambda \Psi(\mathbf{x}),
\label{eq:optimization}
\end{equation}
where $\Psi(\mathbf{x})$ is a regularization term that promotes sparsity by means of the L1-norm of ${\bf x}$ in one of the following standard bases: discrete cosine (DCT) basis~\cite{jafarpour2009transform, gu2009parametric, shi2015fluorescence}, wavelet (WT) basis~\cite{ravishankar2010mr, mallat1999wavelet, lai2016image} and total variation (TV) denoising~\cite{poon2015role, yu2009compressed, block2007undersampled}. $\lambda$ is a scalar (hyperparameter) that strikes the balance between fidelity to the compressed measurements and the sparsity prior. We then use the proximal gradient descent algorithm~\cite {boyd2004convex} to solve the convex optimization problem (\ref{eq:optimization}) by alternating between two steps (Fig.~\ref{fig:1}b):
At iteration $t$, the intermediate estimate $\mathbf{z}_t$ is first computed from $\mathbf{x}_t$ by doing the gradient step:
\begin{equation}
\mathbf{z}_t = \mathbf{x}_t - \alpha_t \nabla_{\mathbf{x}_t} (\left\| \mathcal{A}(\mathbf{x}_t) - \mathbf{y} \right\|_2^2).
\label{eq:optimization_step1}
\end{equation}
The intermediate estimate $\mathbf{z}_t$ is then projected back to the feasible solution space to produce the next iterate $\mathbf{x}_{t+1}$ via the proximal operator $\text{prox}_{\lambda \Psi}(\mathbf{z}_t) : \mathbb{R}^n \to \mathbb{R}^n$:
\begin{equation}
\mathbf{x}_{t+1} = \text{prox}_{\lambda \Psi}(\mathbf{z}_t)= \argmin_{\mathbf{z} \in \mathbb{R}^n} (\Psi({\bf z}_t) + \frac{1}{2\lambda} || \mathbf{x}_t - \mathbf{z}_t ||_2^2).
\label{eq:optimization_step2}
\end{equation}
cryoSENSE efficiently solves Problem (\ref{eq:optimization_step2}) at each iteration through the soft-thresholding operator~\cite{boyd2004convex}. We follow these steps until convergence. 

\subsection{Image Recovery with Generative Prior} 

While sparsity provides a universal prior for image recovery, our second, complementary approach in cryoSENSE makes a far less restrictive assumption, learning a prior directly from data. Here, cryoSENSE assumes that cryo-EM images lie on a low-dimensional manifold $\mathcal{M}$, which can be learned using a DDPM. Our goal is to guide the diffusion process to generate images that are consistent with the measurements $\mathbf{y}$ while preserving the structural fidelity of biomolecular images. 

\noindent \textbf{Denoising Diffusion Probabilistic Models (DDPMs).} cryoSENSE uses a DDPM to approximate $\mathcal{M}$ by learning to reverse a gradual noising process~\cite{ho2020denoisingdiffusionprobabilisticmodels, song2021score}. Given a total time $T$, let $\mathbf{x}_T \sim \mathcal{N}(\mathbf{0}, \mathbf{I})$ be pure Gaussian noise and $\mathbf{x}_0$ be a cryo-EM image. These models define a \emph{forward} process $\{ \mathbf{x}_t \}_{t=0}^T$, where $t \in [0, T]$, that progressively transform $\mathbf{x}_0$ into $\mathbf{x}_T$. The forward process~\cite{song2021score} can be modeled using the following stochastic differential equation (SDE):
\begin{equation}
    d\mathbf{x} = -\frac{\beta_t}{2} \mathbf{x} \, dt + \sqrt{\beta_t} \, d\mathbf{w},
\end{equation}
where $\beta_t : \mathbb{R} \rightarrow \mathbb{R}>0$ is the noise scheduler and $\mathbf{w}$ is the standard Weiner process. To reverse this process and denoise the image, cryoSENSE uses the corresponding reverse-time SDE~\cite{ANDERSON1982313}, known as the \emph{reverse} process:
\begin{equation}
    d\mathbf{x} = \left[ -\frac{\beta_t}{2} \mathbf{x} - \beta_t \nabla_{\mathbf{x}_t} \log p(\mathbf{x}_t) \right] dt + \sqrt{\beta_t} \, d\bar{\mathbf{w}},
\label{eq:reverse_sde}
\end{equation}
where $\bar{\mathbf{w}}$ is the reverse-time Wiener process, and $\nabla_{\mathbf{x}_t} \log p(\mathbf{x}_t)$ is the \textit{score function}. $\nabla_{\mathbf{x}_t} \log p(\mathbf{x}_t)$ is approximated by a neural network $\mathbf{s}_{\theta}(\mathbf{x}_t, t):\mathbb{R}^n\rightarrow \mathbb{R}^n$, which is trained via score matching methods~\cite{Vincent2011}. Sampling from the DDPM involves substituting $s_\theta(\mathbf{x}_t,t)$ into Equation (\ref{eq:reverse_sde}) and solving the reverse-time SDE.

\noindent {\bf Posterior sampling.} To guide the DDPM toward high-quality image reconstructions that agree with measurements $\mathbf{y}$, cryoSENSE modifies the reverse process to sample from the posterior distribution $p(\mathbf{x}_t | \mathbf{y})$ rather than the unconditional data distribution $p(\mathbf{x}_t)$~\cite{chung2023diffusion}. By guiding the diffusion trajectory toward samples that agree with $\mathbf{y}$, cryoSENSE ensures that the generated reconstructions faithfully reflect the information contained in the measurements (Fig.~\ref{fig:1}c). To sample from $p(\mathbf{x}_t | \mathbf{y})$, we use Bayes' rule to modify the reverse process:
\begin{equation}
\nabla_{\mathbf{x}_t} \log p(\mathbf{x}_t|\mathbf{y}) = \nabla_{\mathbf{x}_t} \log p(\mathbf{x}_t) + \nabla_{\mathbf{x}_t} \log p(\mathbf{y}|\mathbf{x}_t).
\label{eq:bayes_rule}
\end{equation}
This enables sampling from the posterior distribution $p(\mathbf{x}_t|\mathbf{y})$  by substituting Equation (\ref{eq:bayes_rule}) into Equation (\ref{eq:reverse_sde}):
\begin{equation}
\begin{aligned}
d\mathbf{x} = {} &
\biggl[ -\frac{\beta_t}{2}\,\mathbf{x}               
      - \beta_t \Bigl( \nabla_{\mathbf{x}_t}\log p(\mathbf{x}_t)  
\\
& \qquad \quad 
      {}+ \nabla_{\mathbf{x}_t}\log p(\mathbf{y}|\mathbf{x}_t)
      \Bigr) \biggr] dt                                
      + \sqrt{\beta_t}\, d\bar{\mathbf{w}} .
\end{aligned}
\label{eq:reverse_sde_posterior}
\end{equation}
This leaves us with two terms in Equation (\ref{eq:reverse_sde_posterior}) to compute. 
The first term, $\nabla_{\mathbf{x}_t} \log p(\mathbf{x}_t)$, can be approximated by $\mathbf{s}_{\theta}(\mathbf{x}_t, t)$, while the second term lacks a closed-form solution as it depends on both $\mathbf{x}_t$ and timestep $t$. To approximate $\nabla_{\mathbf{x}_t} \log p(\mathbf{y}|\mathbf{x}_t)$, cryoSENSE obtains a denoised estimate $\widehat{\mathbf{x}}_0$ of the clean image at each timestep $t$ via Tweedie's formula~\cite{efron2011tweedie, kim2021noise2score}, which expresses the posterior mean as:
\begin{equation}
\widehat{\mathbf{x}}_0 := \mathbb{E}_{\mathbf{x}_t \sim p(\mathbf{x}_t | \mathbf{x}_0)}[\mathbf{x}_0 | \mathbf{x}_t] = \mathbf{x}_t + \sigma_t^2 \nabla_{\mathbf{x}_t} \log p(\mathbf{x}_t),
\label{eq:post_mean_tweedie}
\end{equation}
where $\sigma_t^2$ is a hyperparameter. The denoised estimate $\widehat{\mathbf{x}}_0$ allows us to compute $\nabla_{\mathbf{x}_t} \log p(\mathbf{y}|\widehat{\mathbf{x}}_0)$. Assuming $\mathcal{A}(\mathbf{x}_0)$ is differentiable, this enables us to approximate $\nabla_{\mathbf{x}_t} \log p(\mathbf{y}|\mathbf{x}_t)$ as $\nabla_{\mathbf{x}_t} \log p(\mathbf{y}|\mathbf{x}_t) \simeq \nabla_{\mathbf{x}_t} \log p(\mathbf{y}|\widehat{\mathbf{x}}_{0})$~\cite{chung2023diffusion}. Assuming $\boldsymbol{\eta}$ is Gaussian noise with variance $\sigma^2$, $p(\mathbf{y}|\mathbf{x}_0)$ takes the form of:
\begin{equation}
\begin{aligned}
p(\mathbf{y}|\mathbf{x}_0)
&\sim \mathcal{N}\!\left(
   \mathbf{y}| \mathcal{A}(\mathbf{x}_0),\, \sigma^2\mathbf{I}\right) \\
&= \frac{1}{\sqrt{(2\pi)^n \sigma^{2n}}}
   \exp\!\left(
     -\frac{\|\mathcal{A}(\mathbf{x}_0)-\mathbf{y}\|_2^2}{2\sigma^2}
   \right).
\end{aligned}
\label{eq:likelihood}
\end{equation}
By taking the gradient of the log-likelihood $\log p(\mathbf{y}|\widehat{\mathbf{x}}_0)$ with respect to $\mathbf{x}_t$, we end up with the approximation:
\begin{equation}
\begin{aligned}
\nabla_{\mathbf{x}_t}\log p(\mathbf{y}|\mathbf{x}_t)
&\simeq
\nabla_{\mathbf{x}_t}\log p(\mathbf{y}|\widehat{\mathbf{x}}_{0}) \\
&= -\frac{1}{\sigma^2}\,
\nabla_{\mathbf{x}_t}
  \bigl\|\mathcal{A}(\widehat{\mathbf{x}}_{0})-\mathbf{y}\bigr\|_2^2 .
\end{aligned}
\label{eq:consistency_gradient}
\end{equation}

\noindent This approximation allows cryoSENSE to replace the term $\nabla_{\mathbf{x}_t} \log p(\mathbf{y}|\mathbf{x}_t)$ with $\nabla_{\mathbf{x}_t} \log p(\mathbf{y}|\widehat{\mathbf{x}}_{0})$ and facilitate efficient posterior sampling for cryo-EM image reconstruction of $\mathbf{x}^*$. To realize posterior sampling efficiently over a fixed $T$ number of steps, cryoSENSE integrates Nesterov-accelerated gradients to steer the reverse diffusion process~\cite{wang2025pfdifftrainingfreeaccelerationdiffusion, li2022hessianfreehighresolutionnesterovacceleration, d1b0448b09004d66bc7f81fa21e44678}. For more details on the implementation of cryoSENSE, we refer to the supplemental material.

%% file: sec/3_results.tex
\begin{table*}[t]
    \centering
    \caption{
    Average LPIPS and SSIM scores for cryoSENSE Fourier-space masking reconstructions across different masking strategies and compression factor $C$ (rounded to the nearest tenth).
    }
    \label{tab:fourier_space}
    \resizebox{0.7\textwidth}{!}{%
    \begin{tabular}{l|c|cccc|cccc|cccc}
    \toprule
    \multirow{2}{*}{Prior} & Masking & \multicolumn{4}{c|}{Uniform subsampling} 
    & \multicolumn{4}{c|}{Annular ring} 
    & \multicolumn{4}{c}{Radial spoke} \\
    \cmidrule(lr){2-14}
     & $C$ & 10 & 2.5 & 1.4 & 1 
     & 10 & 2.5 & 1.4 & 1 
     & 10 & 2.5 & 1.4 & 1 \\
    \midrule
    \multirow{2}{*}{Sparse - DCT}
     & LPIPS ($\downarrow$) & 0.21 & 0.14 & 0.05 & \textbf{0.00} & \textbf{0.25} & \textbf{0.13} & 0.05 & \textbf{0.00} & \textbf{0.23} & 0.12 & \textbf{0.04} & \textbf{0.00} \\
     & SSIM ($\uparrow$) & \textbf{0.26} & 0.66 & 0.92 & \textbf{1.00} & \textbf{0.15} & 0.44 & \textbf{0.75} & \textbf{1.00} & \textbf{0.28} & \textbf{0.72} & \textbf{0.93} & \textbf{1.00} \\
    \midrule
    \multirow{2}{*}{Sparse - WT}
     & LPIPS ($\downarrow$) & 0.30 & \textbf{0.12} & \textbf{0.04} & \textbf{0.00} & 0.36 & 0.17 & \textbf{0.05} & \textbf{0.00} & 0.30 & \textbf{0.11} & 0.04 & \textbf{0.00} \\
     & SSIM ($\uparrow$) & 0.26 & \textbf{0.67} & \textbf{0.92} & \textbf{1.00} & 0.15 & 0.42 & 0.74 & \textbf{1.00} & 0.23 & 0.71 & 0.92 & \textbf{1.00} \\
    \midrule
    \multirow{2}{*}{Sparse - TV}
     & LPIPS ($\downarrow$) & 0.35 & 0.29 & 0.13 & \textbf{0.00} & 0.37 & 0.30 & 0.19 & \textbf{0.00} & 0.37 & 0.30 & 0.11 & \textbf{0.00} \\
     & SSIM ($\uparrow$) & 0.07 & 0.35 & 0.81 & \textbf{1.00} & 0.04 & \textbf{0.48} & 0.74 & \textbf{1.00} & 0.07 & 0.37 & 0.91 & \textbf{1.00} \\
    \midrule
    \multirow{2}{*}{Gen - DDPM}
     & LPIPS ($\downarrow$) & \textbf{0.21} & 0.13 & 0.05 & \textbf{0.00} & 0.26 & 0.13 & 0.06 & \textbf{0.00} & 0.20 & 0.11 & 0.04 & \textbf{0.00} \\
     & SSIM ($\uparrow$) & 0.22 & 0.63 & 0.91 & \textbf{1.00} & 0.14 & 0.44 & 0.72 & \textbf{1.00} & 0.23 & 0.70 & 0.92 & \textbf{1.00} \\
    \bottomrule
    \end{tabular}
    }
\end{table*}

\section{Experimental Results}
\label{sec:results}
\vspace{-0.2cm}
We empirically evaluate cryoSENSE's accuracy in recovering images and assess the biological validity of the recovered images for downstream analyses. Section~\ref{results:experimental} details the specific proteins and datasets used and our evaluation metrics.  Section~\ref{sec:reconstruction_fidelity} reports reconstruction fidelity and fundamental acquisition limits toward reconstructing cryo-EM images and 3D volumes. Finally, in Sections~\ref{sec:cryoDRGN} and~\ref{sec:modelangelo} we assess the biological utility of cryoSENSE reconstructions for downstream structural analysis, including conformational heterogeneity identification and atomic model building.


\subsection{Experimental Setup}
\label{results:experimental}
\vspace{-0.2cm}
\textbf{Datasets.} We evaluate cryoSENSE on five publicly available cryo-EM datasets: EMPIAR-10076~\cite{davis2016modular} (ribosomal complex), EMPIAR-10648~\cite{saur2020fragment} (protein-ligand complex), EMPIAR-10166~\cite{haselbach2017long} (protein complex responsible for cellular homeostasis), EMPIAR-11526~\cite{sun2023ksga} (ribosomal subunit in \emph{E. coli}), and EMPIAR-10786~\cite{harris2022selective} (substance P-neurokinin receptor G protein complex). These datasets comprise single-particle cryo-EM images with corresponding Contrast Transfer Function (CTF) and pose metadata. All EMPIAR datasets were collected at 128$\times$128 resolution, with the exception of EMPIAR-10648 experiments, which was collected at 256$\times$256 resolution. Full dataset details and preprocessing protocols are provided in the supplementary material. In the main text, we include results from EMPIAR-10076 and EMPIAR-10648, and leave the remainder of the proteins in the supplemental material.

\noindent \textbf{Evaluation metrics.} To assess image reconstruction quality, we use the Structural Similarity Index Measure (SSIM)~\cite{wang2004image} and Learned Perceptual Image Patch Similarity (LPIPS)~\cite{zhang2018unreasonable}, which provide complementary measures of structural and perceptual similarity between reconstructed and ground-truth images. \textcolor{black}{Peak-signal-to-noise-ratio (PSNR) results are provided in the supplementary material}. For 3D volume evaluation, we backproject images using the provided particle poses to reconstruct three volumes: (1) the original images, (2) cryoSENSE-sparse prior with DCT, which achieved the best overall performance among all sparse priors across our image reconstruction experiments (see supplementary material), and (3) cryoSENSE-generative prior reconstructed images. We then compute the achieved resolution (in \AA) using Fourier Shell Correlation (FSC) at the 0.143 cutoff~\cite{rosenthal2003optimal} with no mask, and assess real-space similarity using the volume-to-volume correlation coefficient (VC) via the \texttt{measure correlation} command in UCSF ChimeraX~\cite{meng2023ucsf}. 

\noindent {\bf Measurement process.} In our experiments, we implement the measurement operator $\mathcal{A}$ through both pixel-space masking and Fourier-space masking. We define the compression factor as $C=n/m \geq 1$, where a higher $C$ corresponds to fewer acquired measurements $m$ taken. For pixel-space masking, we linearly project $\mathbf{x}$ through a set of $b$ binarized random masks. For each masked image, we apply a non-overlapping kernel-wise summation over patches of size $K \times K$ across the entire image. This operation reduces spatial resolution and produces low-resolution measurements $\mathbf{y}$ of dimension $m = bn/K^2$, resulting in a compression factor of $C=K^2/b$. For Fourier-space masking, $\mathcal{A}$ first takes the Fourier transform of $\mathbf{x}$ and then applies a single binary mask to subsample $m$ total Fourier coefficients, such that $m=n/C$. To choose which Fourier coefficients to subsample, we explore three distinct masking strategies: (1) uniform subsampling, (2) annular ring sampling with a low-frequency bias, and (3) radial spoke sampling. For all Fourier-space masking 3D volume reconstructions, we use uniform subsampling, which is the best-performing masking strategy among all experiments. 
Full details about the experiments can be found in the supplemental material.

\subsection{High-fidelity Reconstruction and Fundamental Acquisition Limits }\label{sec:reconstruction_fidelity}

We first benchmark cryoSENSE's ability to reconstruct cryo-EM images from highly compressed measurements on EMPIAR-10076. We evaluate both pixel-space masking, across five downsampling levels $K \in \{ 2, 4, 8, 16, 32\}$ and varying $C$, and Fourier-space masking, varying $C$. \textcolor{black}{The cryo-EM images from EMPIAR-10076 already contain experimental noise. For the main experiments, the compressed measurements are generated from these images without adding any additional measurement noise after applying the measurement operator.} In the supplementary material, we additionally repeat the reconstruction process under additive Gaussian \textcolor{black}{measurement} noise and show that moderate ($\sigma^2 = 0.01$) noise minimally impacts reconstruction quality. \textcolor{black}{We further provide a detailed discussion of the measurement noise model and evaluate noise robustness across different measurement SNR levels.}

Table~\ref{tab:real_space} details reconstruction performance with pixel-space masking. While LPIPS and SSIM both improve as expected with decreasing $C$, generative priors outperform sparse priors at higher compression factors. Notably, generative priors outperform sparse priors across all compression factors at $K=16$ and 32, which could be of separate interest when imaging using inexpensive detectors with lower pixel resolution achievable. When $K \leq 8$, sparse priors match or slightly outperform generative priors. We attribute this behavior to the influence of the learned generative prior, where its performance is constrained by the quality of the learned manifold $\mathcal{M}$. In contrast, sparsity prior-based methods are primarily limited by the information contained in $\mathbf{y}$. For reconstruction performance with Fourier-space masking (Table~\ref{tab:fourier_space}), sparse priors match or slightly outperform generative priors across all compression factors, suggesting that sparse priors are more suitable for Fourier-space masking and moderate compression, while generative priors are more suitable for pixel-space masking under extreme downsampling and compression. Taken together, these results demonstrate that faithful 2D cryo-EM image recovery is possible from highly compressed measurements. Defining an LPIPS of $\simeq$ 0.15 as an empirically acceptable threshold for image quality (which, as shown in the following sections, also enables high-fidelity downstream biological analyses), cryo-EM images can be compressed to a compression factor of $C=2.7$ in pixel-space masking and $C=2.5$ in Fourier-space masking.

\begin{table}[t]
    \centering
    \caption{Reconstruction performance on EMPIAR-10076 with pixel-space and Fourier-space masking,
    benchmarking FSC resolution values (\AA), real-space volume correlations with the ground-truth structure (VC)\textcolor{black}{, and pose estimation accuracy, reporting median angular error (AE, in degrees) and median translational shift error (TSE, in \AA)}. Compression factors $C$ are rounded to the nearest tenth. Ground truth structure has a resolution of 6.65~\AA.
    }
    \label{tab:3d_volumes}
    \resizebox{0.47\textwidth}{!}{%
    \setlength{\tabcolsep}{2pt}
    \begin{tabular}{l|cccccc|ccccc}
    \toprule
    & \multicolumn{6}{c|}{Pixel-space Masking} & \multicolumn{5}{c}{Fourier-space Masking} \\
    \cmidrule(lr){2-7} \cmidrule(lr){8-12}
    Prior & $K$ & $C$ & \AA{} $\downarrow$ & VC $\uparrow$ & \textcolor{black}{AE} & \textcolor{black}{TSE} & $C$ & \AA{} $\downarrow$ & VC $\uparrow$ & \textcolor{black}{AE} & \textcolor{black}{TSE} \\
    \midrule
    \multirow{5}{*}{\makecell{Sparse -\\DCT}}
    & 2 & 4 & \textbf{9.53} & \textbf{0.86} & \textcolor{black}{\textbf{2.6}} & \textcolor{black}{\textbf{1.4}} & 10 & \textbf{10.75} & \textbf{0.87} & \textcolor{black}{130.9} & \textcolor{black}{28.6} \\
    & 4 & 2 & \textbf{7.35} & \textbf{0.91} & \textcolor{black}{\textbf{1.4}} & \textcolor{black}{\textbf{0.8}} & 5 & \textbf{8.38} & \textbf{0.95} & \textcolor{black}{124.2} & \textcolor{black}{22.0} \\
    & 8 & 1.5 & \textbf{6.65} & \textbf{0.95} & \textcolor{black}{\textbf{1.4}} & \textcolor{black}{\textbf{0.7}} & 3.3 & \textbf{7.35} & \textbf{0.97} & \textcolor{black}{92.2} & \textcolor{black}{12.7} \\
    & 16 & 1.3 & 7.35 & 0.82 & \textcolor{black}{2.8} & \textcolor{black}{1.4} & 2.5 & \textbf{6.65} & \textbf{0.98} & \textcolor{black}{4.9} & \textcolor{black}{\textbf{2.3}} \\
    & 32 & 1.4 & 14.46 & 0.76 & \textcolor{black}{129.8} & \textcolor{black}{24.3} & 2 & \textbf{6.65} & \textbf{0.99} & \textcolor{black}{1.9} & \textcolor{black}{\textbf{0.9}} \\
    \cmidrule(lr){1-12}
    \multirow{5}{*}{\makecell{Gen -\\DDPM}}
    & 2 & 4 & 9.98 & 0.83 & \textcolor{black}{3.1} & \textcolor{black}{1.6} & 10 & 11.03 & 0.77 & \textcolor{black}{\textbf{128.8}} & \textcolor{black}{\textbf{22.1}} \\
    & 4 & 2 & 8.38 & 0.80 & \textcolor{black}{1.9} & \textcolor{black}{1.0} & 5 & 9.11 & 0.91 & \textcolor{black}{\textbf{116.4}} & \textcolor{black}{\textbf{16.6}} \\
    & 8 & 1.5 & 8.06 & 0.81 & \textcolor{black}{4.3} & \textcolor{black}{2.2} & 3.3 & 7.62 & 0.96 & \textcolor{black}{\textbf{72.7}} & \textcolor{black}{\textbf{8.8}} \\
    & 16 & 1.3 & \textbf{6.99} & \textbf{0.97} & \textcolor{black}{\textbf{1.7}} & \textcolor{black}{\textbf{0.9}} & 2.5 & 6.76 & 0.98 & \textcolor{black}{\textbf{3.5}} & \textcolor{black}{2.5} \\
    & 32 & 1.4 & \textbf{7.91} & \textbf{0.93} & \textcolor{black}{\textbf{4.6}} & \textcolor{black}{\textbf{2.3}} & 2 & \textbf{6.65} & \textbf{0.99} & \textcolor{black}{\textbf{1.8}} & \textcolor{black}{\textbf{0.9}} \\
    \bottomrule
    \end{tabular}
    }
\end{table}

We additionally assess downstream structural fidelity by reconstructing 3D volumes with sparse priors (DCT) and generative priors (DDPM) using known particle poses (Table~\ref{tab:3d_volumes}). In pixel-space masking, following our 2D analysis, we select $C$ for each $K$ based on achieving an LPIPS threshold of approximately 0.15 with DDPM. DCT notably recovers ground-truth resolution at $C=1.45$, while DDPM provides better reconstructions at large values of $K$. Similar to the 2D image case, in Fourier-space masking, both methods maintain high fidelity at compression factors up to $C=2.5$, with DCT outperforming DDPM across all compression factors. \textcolor{black}{We further evaluate pose estimation robustness by performing homogeneous refinement in CryoSPARC~\cite{punjani2017cryosparc} on cryoSENSE reconstructions (Table~\ref{tab:3d_volumes}). In pixel-space masking, poses remain accurate (median angular error $<5^\circ$, shift error $<2.5$~\AA) up to $K=16$, with DDPM maintaining accurate poses at $K=32$ whereas DCT does not. In Fourier-space masking, poses remain accurate up to $C=2.5$, consistent with the compression limits identified above.} These results highlight the flexibility and robustness of cryoSENSE in masking strategies and compression and provide a powerful strategy for addressing cryo-EM throughput.

\begin{figure*}[t!]
    \centering
    \includegraphics[width=0.85\linewidth]{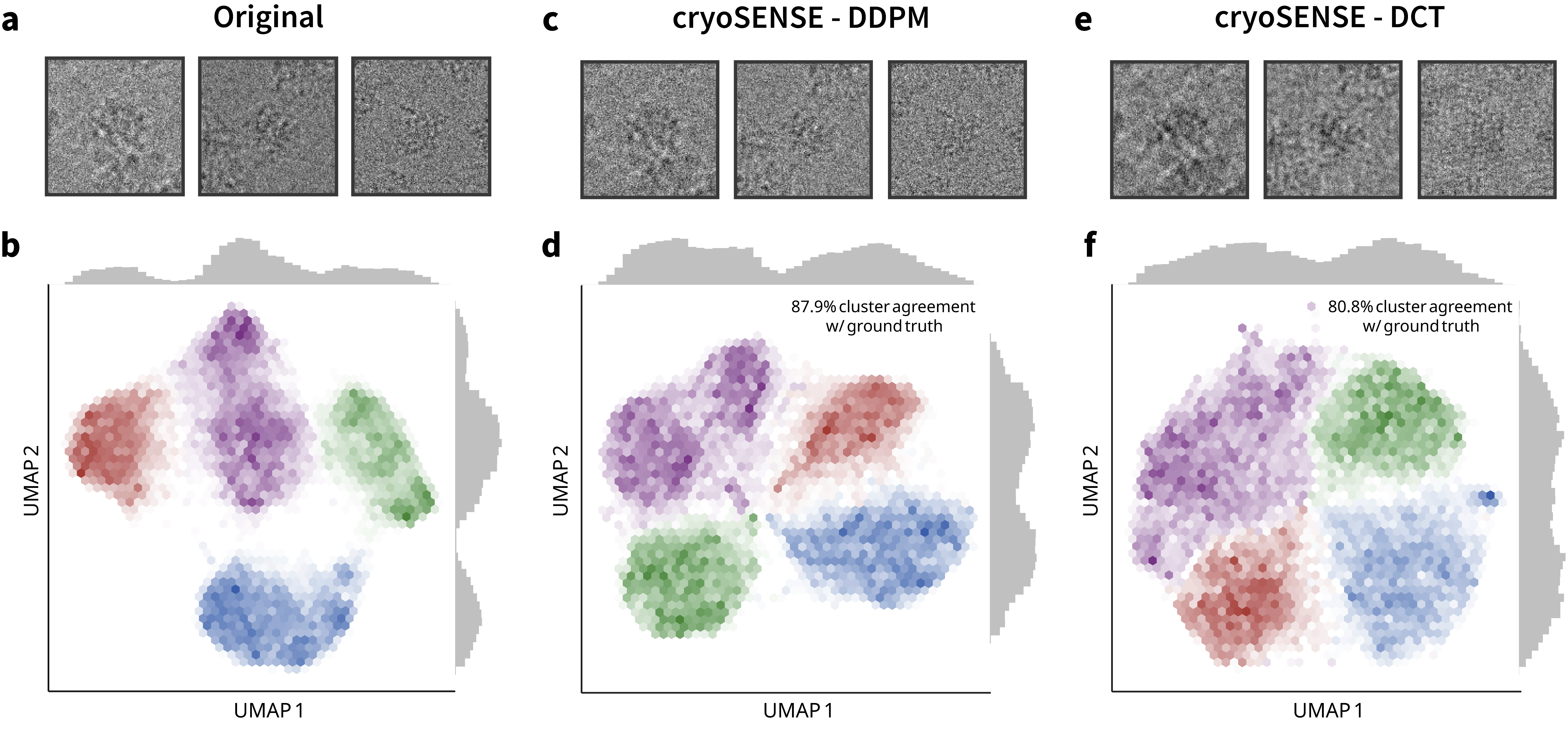}
    \caption{{\textbf{a}, Example EMPIAR-10076 particle images used for CryoDRGN heterogeneity analysis. \textbf{b}, UMAP projection of CryoDRGN latent space trained on original 128$\times$128 images, showing four distinct conformational clusters colored by GMM labels. \textbf{c}, Example cryoSENSE - DDPM reconstructions obtained from $K = 16$ downsampled images with $C=1.25$. \textbf{d}, UMAP projection of CryoDRGN latent space trained on DDPM reconstructions, colored by original GMM cluster labels from \textbf{b}. \textbf{e}, Example cryoSENSE - DCT reconstructions obtained from $K = 16$ downsampled images with $C=1.25$. \textbf{f}, UMAP projection of CryoDRGN latent space trained on DCT reconstructions, colored by original GMM cluster labels from \textbf{b}.}}
    \label{fig:3}
\end{figure*}

\subsection{Conformational Heterogeneity Recovery via CryoDRGN}
\label{sec:cryoDRGN}
We evaluate whether cryoSENSE reconstructions preserve conformational heterogeneity by comparing CryoDRGN~\cite{zhong2021cryodrgn} latent space analysis on original and recovered particle sets. We use EMPIAR-10076 validation images (Fig.~\ref{fig:3}a) at 128$\times$128 resolution and train a CryoDRGN model to embed the particle images into an 8-dimensional latent space. We visualize the latent space using UMAP (Fig.~\ref{fig:3}b) and observe four distinct conformational clusters. Gaussian mixture modeling (GMM) identifies cluster assignments.

We then apply cryoSENSE pixel-space masking to reconstruct the $K=16$ downsampled images with $C=1.25$ using DCT and DDPM priors (Fig.~\ref{fig:3}c and e) and repeat the CryoDRGN workflow. Fig.~\ref{fig:3}d and f show the UMAP projection colored by GMM labels assigned from the original dataset (Fig.~\ref{fig:3}b). To quantify how well cryoSENSE preserves conformational heterogeneity, we compute the fraction of particles in each cryoSENSE cluster that match the corresponding cluster assignment from the original dataset. For DCT reconstructions, cluster agreement ranges from 74.2\% to 89.0\% (80.8\% on average). For DDPM reconstructions, agreement ranges from 82.1\% to 91.4\% (87.9\% on average). These results demonstrate that cryoSENSE enables the recovery of conformational heterogeneity from highly compressed, low-resolution measurements, and that in higher downsampling regimes ($K=16$), the DDPM prior provides superior reconstructions, leading to higher cluster correspondence than DCT. 

\subsection{Atomic Model Recovery via ModelAngelo}
\label{sec:modelangelo}

\begin{figure*}[t!]
\vspace{-0.5cm}
    \centering
    \includegraphics[width=1\linewidth]{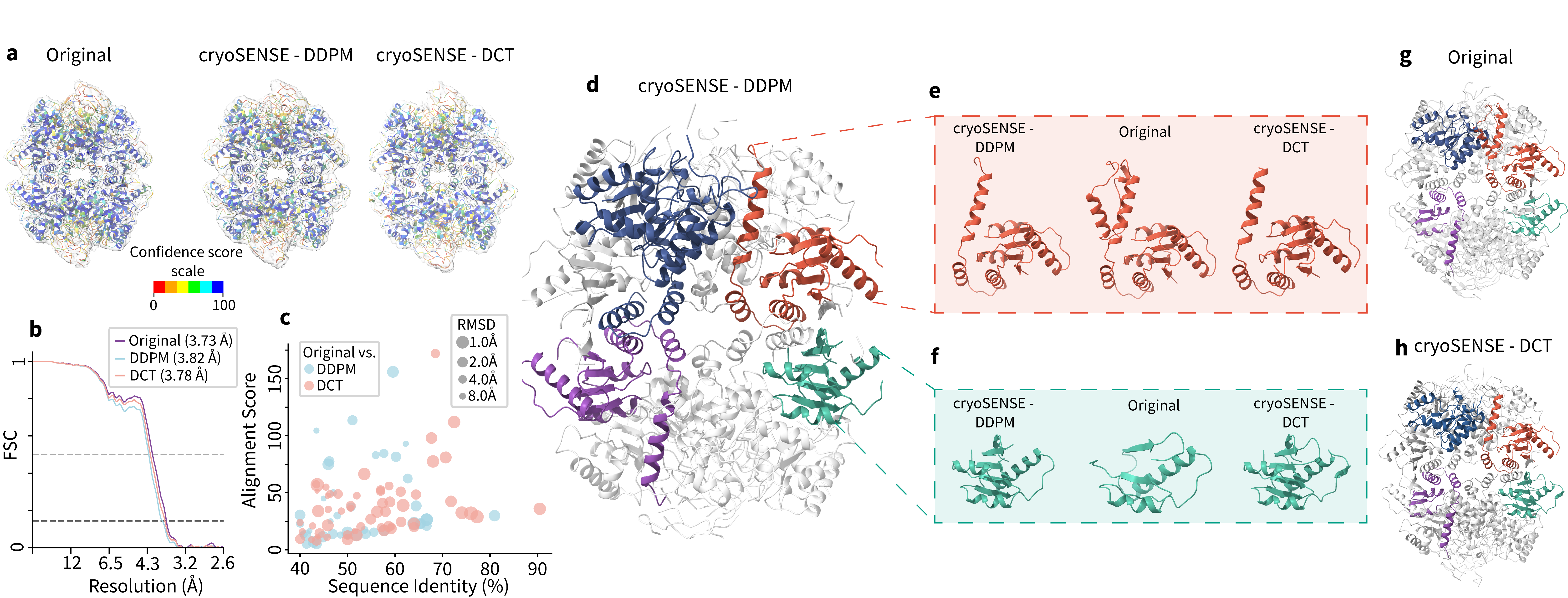}
    \caption{{\textbf{a}, Atomic models from ModelAngelo fit into reconstructed cryo-EM 3D volumes for the Original dataset (EMPIAR-10648) and cryoSENSE reconstructions ($K=2$, $C=1.33$) using DCT and DDPM priors. Chains are colored by ModelAngelo prediction confidence scores. \textbf{b}, FSC curves for the three corresponding cryo-EM 3D volumes. \textbf{c}, Chain-level sequence alignment score vs.\ sequence identity for matched chains between Original vs.\ DDPM and Original vs.\ DCT; marker size reflects backbone RMSD. \textbf{d}, cryoSENSE - DDPM atomic model with four representative chain regions highlighted, corresponding to the same colored regions shown in \textbf{g}. \textbf{e}, First example structural comparison of the highlighted region in the Original model with its corresponding DDPM- and DCT-reconstructed regions. \textbf{f}, Second example structural comparison of the highlighted region in the Original model with its corresponding DDPM- and DCT-reconstructed regions. \textbf{g}, Original atomic model with four chain regions highlighted corresponding to the same regions as in \textbf{d} and \textbf{h}. \textbf{h}, cryoSENSE - DCT atomic model with four representative chain regions highlighted, corresponding to the same colored regions shown in \textbf{g}.}}
    \label{fig:4}
\end{figure*}

To assess the structural accuracy of atomic models obtained from cryoSENSE reconstructions, we analyze atomic structures generated from cryo-EM 3D volumes of EMPIAR-10648 using ModelAngelo~\cite{jamali2024automated} (see supplementary material for inference details). We evaluate three cases: (1) cryo-EM 3D volume reconstructed from the validation particle set (Original) and cryoSENSE pixel-space masking reconstructions after $K=2$ downsampling with $C=1.33$ using (2) DCT and (3) DDPM priors. For all cases, atomic model building is performed using the corresponding cryo-EM 3D volumes and the FASTA sequence of Protein Data Bank (PDB) entry \texttt{6ttf}, representing the sequence associated with EMPIAR-10648. Atomic models for each condition are colored by the ModelAngelo per-residue prediction confidence scores (Fig.~\ref{fig:4}a).  The mean per-residue confidence scores are similarly high for all three conditions, with Original, DCT, and DDPM achieving 67.3 $\pm$ 33.5, 64.71 $\pm$ 33.82, 62.5 $\pm$ 33.4, respectively. We next examine the resolutions of the underlying 3D maps using FSC (Fig.~\ref{fig:4}b). The three volumes achieve FSC (tight mask) resolutions of 3.73~\AA{} (Original), 3.78~\AA{} (DCT), and 3.82~\AA{} (DDPM), indicating that cryoSENSE retains sufficient structural detail to support reliable atomic model building, with the DCT prior providing the marginally higher-resolution map.

We next assess chain-level structural correspondence between models by performing pairwise alignments and plotting alignment score versus sequence identity (Fig.~\ref{fig:4}c). Polypeptide chains with at least 20 residues are extracted from each atomic model and matched to the Original model using global sequence alignment implemented in Biopython’s \texttt{pairwise2} module, with a gap penalty scoring scheme (match: 2, mismatch: –1, gap open: –2, gap extend: –0.5). Alignments with sequence identity $\geq$ 40\% are retained for structural comparison. Backbone RMSD values are computed from backbone atoms (N, C$\alpha$, C, O) after structural superposition. Across all alignments, comparison to the Original model yields 42 matched chain pairs for DDPM and 58 for DCT, with mean backbone RMSDs (lower is better) of 2.34~\AA\ and 2.07~\AA\ and mean alignment scores (higher is better) of 38.1 and 38.4, respectively, indicating that the DCT prior produces more complete and slightly more accurate chain-level structural recovery than the DDPM prior at $K=2$ downsampling with $C=1.33$.

To illustrate individual chain-level correspondences, we examine representative chain-to-chain alignments shown in Fig.~\ref{fig:4}d–h. Both DDPM and DCT reconstructions recover long chain segments with high sequence identity and strong alignment scores. Among the DDPM matches, the best examples include Original chain \texttt{Aa} aligned to DDPM chains \texttt{Ah} and \texttt{Ak} (0.60 and 0.51 identity; scores 156.0 and 113.5), and chain \texttt{AT} aligned to \texttt{Af} (0.43 identity; score 104.5). The strongest DCT correspondences include Original \texttt{AT} aligned to \texttt{BT} (0.68 identity; score 172.0) and Original \texttt{AU} aligned to \texttt{Ak} (0.72 identity; score 112.0), both involving long chains (96–151 residues).

Together, these results show that cryoSENSE supports accurate atomic model recovery from compressed measurements. At $K=2$ downsampling with $C=1.33$, the DCT prior performs slightly better overall, achieving a higher map resolution (3.78~\AA{} vs.\ 3.82~\AA{}), higher mean confidence scores (64.71 vs.\ 62.5), and more matched chains (58 vs.\ 42) than the DDPM prior. \textcolor{black}{In the supplementary material, we additionally compare cryoSENSE against DMPlug~\cite{wang2024dmplug}, a diffusion-based super-resolution baseline that does not enforce measurement consistency with compressed measurements, and show that cryoSENSE outperforms it across all downstream tasks.}

\vspace{-0.2cm}

\section{Discussion}

\vspace{-0.2cm}
\noindent {\bf Implication for dose fractionation.} In cryo-EM, imaging resolution improves when direct detectors fractionate the total electron dose into a time series of images that can be individually corrected for beam-induced motion~\cite{li2013electron, zheng2017motioncor2, guo2020electron}. Increasing dose fractionation, and hence achieving finer temporal resolution, further enhances attainable spatial resolution by better capturing drift and radiation-damage dynamics~\cite{li2013electron,grant2015measuring, kong2025unbend}. cryoSENSE builds on this principle by using compressed data acquisition to enable higher frame rates under fixed bandwidth and storage limits, and recovering the original images post-acquisition. By recording more finely fractionated frames without data bottlenecks, cryoSENSE preserves the temporal information needed for accurate motion correction and dose weighting, extending these benefits to high-throughput cryo-EM. We further anticipate that a similar hardware–software co-designed framework could drastically enhance temporal resolution in time-resolved in situ electron microscopy~\cite{ross2015opportunities, Jamali2020-2, alcorn2023time, shabeeb2025learning}, where an inherent trade-off between spatial and temporal resolution currently limits dynamic imaging~\cite{de2019resolution}.

\noindent {\bf Limitations and future work.} Our work introduces the first computational demonstration of compressive in situ acquisition that can substantially increase data throughput in cryo-EM imaging of proteins. Here, we note some limitations that open avenues for future research. In our current workflow, masking patterns are directly applied onto single-particle images, whereas, on an electron microscope, acquisition occurs at the micrograph level and thus masking patterns apply to the entire micrograph before particle picking isolates the single-particle images. We expect cryoSENSE to seamlessly extend to full micrograph acquisition, enabling full image recovery across broader fields of view. This would require scaling the DDPM to generate images at the full micrograph scale. In the future, the results from this work will inform the design and integration of hardware~\cite{aharon2006k,aghazadeh2018insense} (e.g., programmable coded apertures and beam shaping elements) with electron microscopes, enabling the direct collection of compressed measurements during imaging.

\noindent {\bf Conclusion.} cryoSENSE introduces a \textcolor{black}{computational realization of a} hardware-software co-designed framework for high-throughput cryo-EM acquisition. The framework is general and supports both handcrafted sparsity priors and learned generative priors: the former enables reconstruction when the sample is unknown or data are limited, while the latter leverages large-scale electron microscopy datasets~\cite{iudin2023empiar} to enhance fidelity when prior information exists. This combined capability allows cryoSENSE to generalize across diverse specimen types, from protein macromolecules to inorganic nanoparticles. Beyond improving cryo-EM throughput, this approach lays the foundation for future programmable detectors that collect compressed measurements directly, transforming high-resolution electron microscopy into a real-time, high spatiotemporal acquisition paradigm for dynamic imaging.

%% file: Appendix/supp.tex





\setcounter{equation}{0}
\renewcommand{\thesection}{\Alph{section}} 
\renewcommand{\thesubsection}{\thesection.\arabic{subsection}}
\renewcommand{\thefigure}{S-\arabic{figure}}
\renewcommand{\thetable}{S-\arabic{table}}


\section{Implementation of Sparse Priors}
\label{appendix:sparse_priors}

We selected three priors in cryoSENSE: sparse L1 recovery with the DCT, sparse L1 recovery with the wavelet transform, and TV regularization. We provide an overview of these methods here.

\noindent \textbf{Sparse DCT recovery}. Sparse DCT recovery assumes that images are sparse in the DCT. We seek to minimize:
\begin{equation*}
    \mathbf{\widehat{x}} = \arg\min_{\mathbf{x}} \left\| \mathbf{y}-\mathcal{A}(\mathbf{x}) \right\|_2^2 + \lambda \, \left\| \Psi_{\text{DCT}} (\mathbf{x}) \right\|_1,
\end{equation*}
where $\Psi_{\text{DCT}}(\mathbf{x})$ denotes the DCT of $\mathbf{x}$. The L1 norm promotes sparsity in the DCT. 

\noindent \textbf{Sparse Wavelet recovery}. Similarly, sparse wavelet reconstruction assumes that images are sparse in the wavelet transform. We solve the following optimization problem:
\begin{equation*}
\mathbf{\widehat{x}} = \arg\min_{\mathbf{x}} \quad \left\| \mathbf{y} - \mathcal{A}(\mathbf{x}) \right\|_2^2 + \lambda \, \left\| \Psi_{\text{WT}} (\mathbf{x}) \right\|_1,
\end{equation*}
where $\Psi_{\text{WT}} (\mathbf{x})$ denotes the wavelet transform of $\mathbf{x}$.

\noindent \textbf{Total Variation regularization}. TV regularization promotes piecewise smoothness by penalizing the gradient magnitude across the image. We minimize:
\begin{equation*}
\mathbf{\widehat{x}} = \arg\min_{\mathbf{x}} \quad \left\| \mathbf{y} - \mathcal{A}(\mathbf{x}) \right\|_2^2 + \lambda \, \mathrm{TV}(\mathbf{x}).
\end{equation*}
Here, $\mathrm{TV}(\mathbf{x})$ is the anisotropic total variation operator over a vector $\mathbf{x}$:
\begin{equation*}
    \mathrm{TV}(\mathbf{x}) = \sum_{i,j} (|x_{i+1,j} - x_{i,j}| + |x_{i,j+1} - x_{i,j}|).
\end{equation*}

To implement the proximal gradient algorithm in sparse DCT reconstruction and sparse wavelet reconstruction, we utilize the Iterative Soft Thresholding Algorithm (ISTA)~\cite{daubechies2004iterative}. At each epoch, we perform a gradient descent step $\mathbf{z}_{t} = \mathbf{x}_t - \alpha \cdot \nabla \|\mathbf{y} - \mathcal{A}(\mathbf{x}_t)\|_2^2$ on the data fidelity term, then apply the proximal operator by computing the transform $\Psi (\mathbf{z}_t)$, applying soft thresholding with threshold $\lambda$, and computing the inverse transform to obtain $\mathbf{x}_{t+1}$. To implement the proximal gradient algorithm in TV regularization, we utilize the $prox\_tv$ package (\href{https://github.com/albarji/proxTV}{https://github.com/albarji/proxTV})~\cite{barbero2011fast, barbero2018modular}.

For sparse DCT reconstruction, sparse wavelet reconstruction, and TV regularization, we solve each optimization problem by performing proximal gradient descent over 200 epochs. There are two hyperparameters to consider: $\lambda$, which governs the strength of the regularizer, and the learning rate (which we set as a constant rate $\alpha$, such that $\alpha_t = \alpha$). To determine these hyperparameters, for each protein, we perform a grid search of $\lambda = [0.1, 0.01, 0.001, 0.0001]$ and $\alpha = [0.001, 0.01, 0.1, 0.5, 1]$ over two training images.

\section{Implementation of Generative Priors}
\label{appendix:generative_priors}

\subsection{Nesterov Momentum Acceleration}
\begin{algorithm}[H]
\caption{\textbf{cryoSENSE sampling with Nesterov momentum.}}
\label{alg:cryogen}
\begin{algorithmic}[1]
\Require $\mathbf{y}$, $\mathcal{A}$, $s_\theta(\mathbf{x}_t, t)$, $\{\alpha_t\}_{t=1}^{T}$, $\{\kappa_t\}_{t=1}^{T}$, $\{\zeta_t\}_{t=1}^{T}$
\State $\mathbf{x}_{T}\sim\mathcal{N}(\mathbf{0},\mathbf{I})$, \; $\mathbf{m}_{T}\leftarrow\mathbf{0}$
\For{$t=T,\dots,1$}
  \State $\mathbf{s} \leftarrow s_\theta(\mathbf{x}_t, t)$
  \State $\widehat{\mathbf{x}}_{0}
  \leftarrow \frac{1}{\sqrt{\bar{\alpha}_t}}
  \bigl(\mathbf{x}_t + (1 - \bar{\alpha}_t)\mathbf{s}\bigr)$
  \State $\mathbf{x}'_{t-1}
  \leftarrow
  \frac{\sqrt{\alpha_t}(1 - \bar{\alpha}_{t-1})}{1 - \bar{\alpha}_t}\mathbf{x}_t
  + \frac{\sqrt{\bar{\alpha}_{t-1}}\beta_t}{1 - \bar{\alpha}_t}\widehat{\mathbf{x}}_0
  + \Tilde{\sigma_t}\mathbf{z}$
  \State $\mathbf{p}_t \leftarrow \widehat{\mathbf{x}}_{0}-\kappa_t\mathbf{m}_t$
  \State $\mathbf{g}_t \leftarrow
  \nabla_{\mathbf{p}_t}\bigl\lVert\mathbf{y}-\mathcal{A}(\mathbf{p}_t)\bigr\rVert_2^2$
  \State $\mathbf{m}_{t-1}\leftarrow\kappa_t\mathbf{m}_t+\zeta_t\mathbf{g}_t$
  \State $\mathbf{x}_{t-1}\leftarrow\mathbf{x}_{t-1}^{'}-\mathbf{m}_{t-1}$
\EndFor
\State \Return $\mathbf{x}_{0}$
\end{algorithmic}
\end{algorithm}

cryoSENSE integrates an accelerated correction step into the standard DDPM sampling procedure. Rather than directly modifying the reverse SDE, we thus apply measurement consistency guidance at each denoising step using a momentum-based approach inspired by Nesterov acceleration. At each diffusion step $t$, we first compute the denoised estimate $\widehat{\mathbf{x}}_0$ of the clean underlying image:
\begin{equation}
\widehat{\mathbf{x}}_0 = \frac{1}{\sqrt{\bar{\alpha}_t}} \left( \mathbf{x}_{t} + (1-\bar{\alpha}_t) s_\theta(\mathbf{x}_t, t) \right),
\end{equation}
where $\alpha_t = 1 - \beta_t$ and $\bar{\alpha}_t = \prod_{s=1}^{t} \alpha_s$. Using $\widehat{\mathbf{x}}_0$, we obtain the unconstrained image $\mathbf{x}_{t-1}'$, which represents the next sample generated by the DDPM without guidance:
\begin{equation}
\mathbf{x}_{t-1}' = \frac{\sqrt{\alpha_t}(1 - \bar{\alpha}_{t-1})}{1 - \bar{\alpha}_t} \mathbf{x}_t 
+ \frac{\sqrt{\bar{\alpha}_{t-1}}\beta_t}{1 - \bar{\alpha}_t} \widehat{\mathbf{x}}_0 
+ \tilde{\sigma_t} \mathbf{z},
\label{eq:discrete_sde}
\end{equation}
where $\mathbf{z} \sim \mathcal{N}(\mathbf{0}, \mathbf{I})$. The reverse diffusion variance $\tilde{\sigma}_t^{2}$ is computed as $
\tilde{\sigma}_t^{2} = \frac{1 - \bar{\alpha}_{t-1}}{1 - \bar{\alpha}_{t}}\,\beta_t
$, as in \cite{ho2020denoisingdiffusionprobabilisticmodels}. Equation (\ref{eq:discrete_sde}) is a natural discretization of the reverse process given a fixed number of steps. To guide the denoising trajectory, cryoSENSE introduces the look-ahead point $\mathbf{p}_t$ by extrapolating from $\widehat{\mathbf{x}}_0$ along the current momentum direction:
\begin{equation}
\mathbf{p}_t = \widehat{\mathbf{x}}_0 - \kappa_t \mathbf{m}_t,
\end{equation}
where $\mathbf{m}_t$ is the accumulated momentum vector and $\kappa_t$ is a time-dependent extrapolation coefficient. The gradient of $\mathbf{p}_t$, denoted as $\mathbf{g}_t = \nabla_{\mathbf{p}_t}
   \lVert\mathbf{y}-\mathcal{A}(\mathbf{p}_t)\rVert_{2}^{2}$, is used to enforce measurement consistency when guiding DDPM toward the corrected image $\mathbf{x}_{t-1}$. At each timestep, we update $\mathbf{m}_t$ to $\mathbf{m}_{t-1}$ via $\mathbf{m}_{t-1} = \kappa_t\mathbf{m}_t + \zeta_t\mathbf{g}_t$, where $\zeta_t$ governs the strength of measurement consistency. We then obtain $\mathbf{x}_{t-1}$ as $\mathbf{x}_{t-1} = \mathbf{x}_{t-1}' - \mathbf{m}_{t-1}$.

\subsection{DDPM Training Details}
\label{appendix:ddpm_training}
The DDPM model \cite{ho2020denoisingdiffusionprobabilisticmodels} was trained within the standard denoising diffusion probabilistic modeling framework using a UNet2D architecture. The network comprises six down-sampling blocks with output channels of $[128, 128, 256, 256, 512, 512]$, mirrored by six up-sampling blocks; self-attention is introduced in the fifth down block (\textit{AttnDownBlock2D}) and the second up block (\textit{AttnUpBlock2D}). Because cryo-EM images are grayscale, both the input and output channel dimensions are set to $1$. Optimization is performed with AdamW ($\beta_1 = 0.95$, $\beta_2 = 0.999$, weight decay $1\times10^{-6}$, Adam epsilon $\epsilon = 1\times10^{-8}$) combined with a cosine learning-rate schedule preceded by $500$ warm-up steps. Training runs on eight NVIDIA A6000 GPUs in mixed-precision (bfloat16), using a per-GPU batch size shown in the table below and gradient-accumulation of $2$. Gradients are clipped to an $\ell_2$-norm of $1.0$, and an exponential moving average of the model parameters is maintained (inverse gamma = $1.0$, power = $0.75$, maximum decay = $0.9999$). Input images are normalized to $[-1, 1]$ and corrupted with a linear noise ($\beta$) schedule across $1000$ diffusion timesteps. The model is trained for $100$ epochs.

Table \ref{tab:training-details} lists the key hyper-parameters for each dataset. The per-GPU batch size was set to the maximum that fits in memory on eight NVIDIA A6000 cards—128 images for 128$\times$128 inputs and 16 images for 256$\times$256 inputs. We used a learning rate of $2\times10^{-4}$ for the 128$\times$128 resolution dataset batches and reduced it by one order of magnitude ($2\times10^{-5}$) for the 256$\times$256 image dataset, following settings similar to those reported in the original DDPM paper \cite{ho2020denoisingdiffusionprobabilisticmodels}.

\begin{table}[htbp]
  \centering
  \caption{Training details for DDPM models on protein datasets.}
  \small 
  \begin{tabular}{lcccc}
    \toprule
    \textbf{Dataset} & \textbf{Resolution} & \textbf{Train Size} & \textbf{LR} & \textbf{Batch} \\
    \midrule
    EMPIAR-10076 & 128$\times$128 & 105,519 & 2e-4 & 128 \\
    EMPIAR-11526 & 128$\times$128 & 200,260 & 2e-4 & 128 \\
    EMPIAR-10166 & 128$\times$128 & 190,904 & 2e-4 & 128 \\
    EMPIAR-10786 & 128$\times$128 & 230,927 & 2e-4 & 128 \\
    EMPIAR-10648 & 256$\times$256 & 187,964 & 2e-5 & 16 \\
    \bottomrule
  \end{tabular}
  \label{tab:training-details}
\end{table}

\subsection{Implementation of Posterior Sampling}

For stability and to ensure that generated images remain within a valid range, we clip both the predicted clean image $\widehat{\mathbf{x}}_{0}$ and the subsequent sample $\mathbf{x}_{t-1}$ to the interval $[-1, 1]$.


        

We additionally set $\zeta_{\min} = 10^{-10}$ to stabilize optimization in the early stages of sampling, ensuring that the influence of the measurement consistency gradient is initially minimal and gradually increases over time. Empirically, we found that setting $\zeta_{\max} = 1.0$ yields stable reconstructions — measured by LPIPS, SSIM, and PSNR — for kernel sizes 2, 4, 8, and 16 (see Section~\ref{appendix:obtaining_linear_measurements} for definition of kernel size). For the largest kernel size 32, we observed that $\zeta_{\max} = 10.0$ provided the best reconstruction quality. Intuitively, lower-resolution measurements required stronger guidance from the measurement consistency gradient to achieve accurate reconstructions.

The momentum coefficients $\kappa_t$ were set linearly between $\kappa_{\min} = 0.1$ and $\kappa_{\max} = 0.9$, following common practice in Nesterov-accelerated gradient methods.

The momentum coefficients, $\zeta_{t}$ and $\kappa_{t}$, vary linearly with timestep $t$ as follows:
\[
\zeta_{t} = \zeta_{\min} + \frac{t - 1}{T - 1}\left(\zeta_{\max} - \zeta_{\min}\right),
\]
\[
\quad \\
\kappa_{t} = \kappa_{\min} + \frac{t - 1}{T - 1}\left(\kappa_{\max} - \kappa_{\min}\right).
\]

\section{Cryo-EM Dataset Details}
\label{app:datasets}

We provide full details on dataset sizes, preprocessing, and resolution settings used in our experiments.

\noindent\textbf{EMPIAR-10166.} Human 26S proteasome bound to Oprozomib~\cite{haselbach2017long}. We use particle images and pose metadata provided by the CESPED benchmark~\cite{sanchez-garcia2024cesped}, with 190,904 training and 23,863 validation particles. Images were originally 284 $\times$ 284 pixels and downsampled to 128 $\times$ 128 using Fourier cropping via the CryoDRGN downsample utility.

\noindent \textbf{EMPIAR-10076.} E. coli large ribosomal subunit assembly intermediates~\cite{davis2016modular}. We used the dataset provided by the CryoDRGN Zenodo repository~\cite{zhong2021cryodrgn}, with 105,519 training particles and 26,380 validation particles. The original images were 320 $\times$ 320 pixels and were downsampled to 128 $\times$ 128 using Fourier cropping for all experiments.

\noindent \textbf{EMPIAR-10648.} PKM2 protein bound to a small-molecule inhibitor~\cite{saur2020fragment}. We used particle images and pose metadata from the CESPED benchmark~\cite{sanchez-garcia2024cesped}, with 187,964 particles for training and 23,496 for validation. Images were provided at 222 $\times$ 222 pixels and were upsampled to 256 $\times$ 256 via bicubic interpolation prior to model input to match the dimensional requirements of our DDPM framework. We verified that this upsampling had no adverse effect on the structural information, as comparable 3D volume resolutions were obtained before and after upsampling.

\noindent \textbf{EMPIAR-10786.} Substance P–Neurokinin Receptor G protein complexes~\cite{harris2022selective}. We used the particle images and pose metadata provided by the CESPED benchmark~\cite{sanchez-garcia2024cesped}, with 230,927 training particles and 28,866 validation particles. Images were downsampled from 184 $\times$ 184 to 128 $\times$ 128 via Fourier cropping.

\noindent \textbf{EMPIAR-11526.} Small ribosomal subunit assembly intermediates in E. coli. We used the dataset released by Sun et al. \cite{sun2023ksga} via Zenodo, with 200,260 training particles and 25,032 validation particles. Images were provided at 256 $\times$ 256 pixels and were downsampled to 128 $\times$ 128 via Fourier cropping.

\section{Obtaining Linear Measurements}
\label{appendix:obtaining_linear_measurements}

\begin{figure*}[t]
  \centering
  \begin{tikzpicture}[
    node distance=0.8cm,
    box/.style={draw, rounded corners=1pt, fill=gray!5, minimum width=4cm, minimum height=2.8cm, align=center, font=\small},
    infobox/.style={draw, rounded corners=1pt, fill=gray!10, minimum width=4cm, minimum height=1.5cm, align=center},
    resultbox/.style={draw, rounded corners=1pt, fill=white, minimum width=4cm, align=center},
    arrow/.style={->, >=stealth, thick},
    scale=0.85
    ]
    
    \node[box] (original) at (1,0) {\includegraphics[width=2cm]{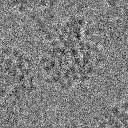}\\[0.2cm] $\mathbf{x}^*$ $(128 \times 128)$};
    \node[box] (mask) at (8,0) {\includegraphics[width=2cm]{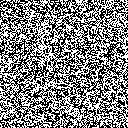}\\[0.2cm] $B_0$};
    
    \node[infobox] (operator) at (4,-3) {$\mathcal{A}$};
    
    \draw[arrow] (original) -- (operator);
    \draw[arrow] (mask) -- (operator);
    
    \def\outputX{9.5}
    
    \foreach \k/\res/\y in {
      2/64/-5.5,
      4/32/-8.5,
      8/16/-11.5,
      16/8/-14.5,
      32/4/-17.5
    } {
      \node[infobox] (k\k) at (0,\y) {$K = \k$\\($\res \times \res$)};
    }
    
    \node[resultbox] (y2) at (\outputX,-5.5) {\includegraphics[width=1.5cm]{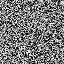}\\$\mathbf{y}$};
    \node[resultbox] (y4) at (\outputX,-8.5) {\includegraphics[width=1.5cm]{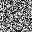}\\$\mathbf{y}$};
    \node[resultbox] (y8) at (\outputX,-11.5) {\includegraphics[width=1.5cm]{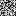}\\$\mathbf{y}$};
    \node[resultbox] (y16) at (\outputX,-14.5) {\includegraphics[width=1.5cm]{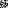}\\$\mathbf{y}$};
    \node[resultbox] (y32) at (\outputX,-17.5) {\includegraphics[width=1.5cm]{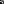}\\$\mathbf{y}$};

    \foreach \k in {2,4,8,16,32} {
      \draw[arrow] (k\k) -- (y\k);
    }
      
  \end{tikzpicture}
  \caption{Visualization of forward operator $\mathcal{A}$ with pixel-space masking over a single mask. The process begins with a $128 \times 128$ input image $\mathbf{x}_0$ and a random binary mask $B_0$. Non-overlapping kernel-wise convolution is applied over $K \times K$ patches, resulting in progressively downsampled measurement resolutions as kernel size increases.}
  \label{app:forward_operator}
\end{figure*}

\noindent \textbf{Pixel-space masking.} In pixel-space masking, the forward operator $\mathcal{A}$ simulates the cryo-EM measurement process by applying structured masking and downsampling to high-resolution images. Given an input image $\mathbf{x}^* \in \mathbb{R}^n$, $\mathcal{A}$ is defined as follows:
\begin{enumerate}
\item Multiple random binary masks $\{ B_i \}_{i=0}^b \in \mathbb{R}^n$, each drawn independently from a Bernoulli distribution with probability $p = 0.5$, are generated and element-wise multiplied with $\mathbf{x}^*$ to simulate partial observations. Each mask $B_i$ randomly selects a subset of pixels to retain.
\item After masking, a kernel-wise summation pooling operation with kernel size $K$ is applied to the masked images, reducing the spatial resolution. This pooling step aggregates pixel values over non-overlapping $K \times K$ blocks, simulating low-resolution measurements.
\item The final measurements $\mathbf{y} \in \mathbb{R}^m$, with $m= bn / K^2$, are collected across all masks.
\end{enumerate}
Formally, the measurement operator can be expressed as:
\[
\mathcal{A}(\mathbf{x}^*) = \left\{ \text{Pool}_K\left(B_i \odot \mathbf{x}^*\right) \right\}_{i=1}^b,
\]
where $\odot$ denotes element-wise multiplication and $\text{Pool}_K$ denotes block-wise summation pooling over kernels of size $K \times K$. Fig.~\ref{app:forward_operator} illustrates the forward operator $\mathcal{A}$ applied with a single random binary mask, demonstrating various pooling kernel sizes $K \times K$ and their corresponding measurement outputs $\mathbf{y}$.

\begin{figure*}[ht]
\centering
\begin{tikzpicture}[
    node distance=1.5cm, 
    box/.style={draw, rounded corners=1pt, fill=gray!5, minimum width=3cm, minimum height=2.8cm, align=center, font=\small},
    opbox/.style={draw, rounded corners=1pt, fill=gray!10, minimum width=2.5cm, minimum height=1cm, align=center, font=\small}, 
    arrow/.style={->, >=stealth, thick},
    operator_label/.style={midway, fill=white, inner sep=1pt, font=\small}, 
    scale=0.85
  ]
    \node[box] (x0_spatial) at (0,0) {\includegraphics[width=2cm]{Appendix/original_image.png}\\[0.2cm] $\mathbf{x}^*$};
  
    \node[box] (x0_fourier) [right=of x0_spatial] {\includegraphics[width=2cm]{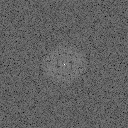}\\[0.2cm] $\mathcal{F}(\mathbf{x}^*)$};
    \draw[arrow] (x0_spatial) -- node[operator_label] {$\mathcal{F}$} (x0_fourier);
  
    \node[opbox] (mask) [above=0.8cm of x0_fourier] {\includegraphics[width=1.5cm]{Appendix/mask_0.png}\\[0.2cm] $B$};
  
    \node[box] (y_fourier) [right=of x0_fourier] {\includegraphics[width=2cm]{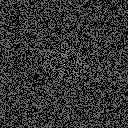}\\[0.2cm] $B \odot \mathcal{F}(\mathbf{x}^*)$}; 
    \draw[arrow] (x0_fourier) -- node[operator_label, anchor=south, xshift=-0.2cm] {} (y_fourier);
    \draw[arrow] (mask) -- (y_fourier); 
  
    \node[box] (y_spatial) [right=of y_fourier] {\includegraphics[width=2cm]{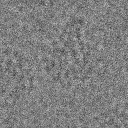}\\[0.2cm] $\mathbf{y}$};
    \draw[arrow] (y_fourier) -- node[operator_label] {$\mathcal{F}^{-1}$} (y_spatial);

\end{tikzpicture}
  \caption{Visualization of forward operator $\mathcal{A}$ with Fourier-space masking. An input image $\mathbf{x}^*$ is transformed into the Fourier domain, $\mathcal{F}(\mathbf{x}^*)$. A binary mask is then applied to the Fourier domain to subsample a subset of Fourier coefficients. The inverse Fourier transform ($\mathcal{F}^{-1}$) yields the corresponding low-resolution measurement $\mathbf{y}$ in the spatial domain.}
  \label{app:fourier_operator}
\end{figure*}

\begin{figure*}[t!]
    \centering
    \includegraphics[width=1\linewidth]{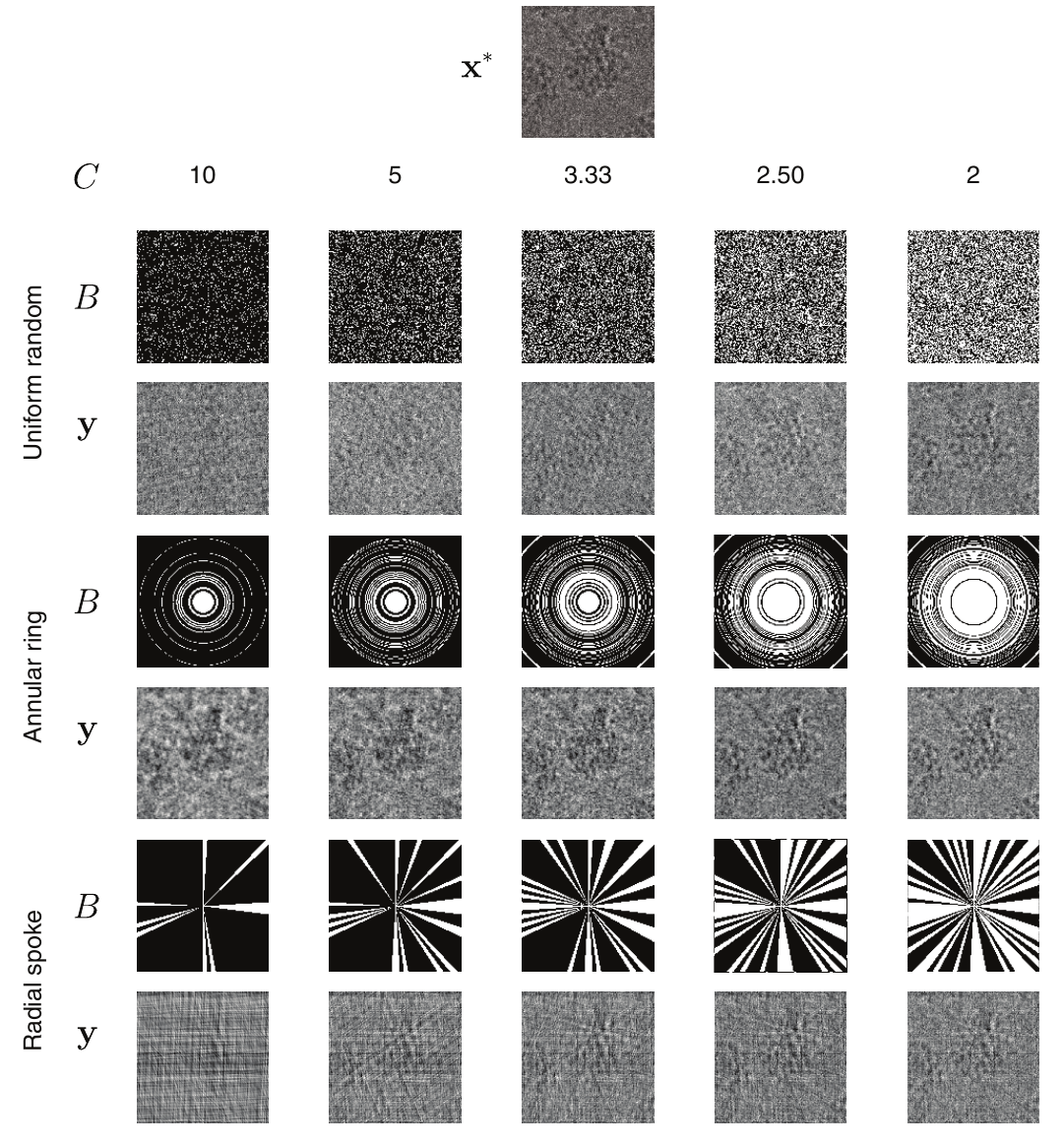}
    \caption{Visualization of uniform, annular ring, and radial spoke Fourier masks at different levels of $C$, and their respective outputs $\mathbf{y}$ given $\mathbf{x}^*$.}
    \label{app:fourier_masks}
\end{figure*}

\noindent \textbf{Fourier-space masking.} In Fourier-space masking, $\mathcal{A}$ is defined as follows (see Fig.~\ref{app:fourier_operator}):
\begin{enumerate}
\item A single binary mask $B \in \mathbb{R}^n$ is generated and element-wise multiplied with the Fourier transform of $\mathbf{x}^*$, denoted as $\mathcal{F}(\mathbf{x}^*)$, to subsample a total of $m$ Fourier coefficients. 
\item The rest of the Fourier coefficients not subsampled are set to zero. The inverse Fourier transform, denoted as $\mathcal{F}^{-1}$, is then applied on the subsampled image to obtain the low-resolution measurement $\mathbf{y}$.
\end{enumerate}
Formally, the measurement operator can be expressed as: 
\[
\mathcal{A}(\mathbf{x}^*) = \mathcal{F}^{-1}(B \odot \mathcal{F}(\mathbf{x}^*)).
\]
In cryoSENSE, we test the following three Fourier masks: (1) uniform, (2) annular ring with low-frequency bias, and (3) radial spoke. These three masks are defined as follows:
\begin{itemize}
\item \textbf{Uniform.} This mask samples $1/C$ many Fourier coefficients, chosen uniformly at random.
\item \textbf{Annular ring with low-frequency bias.} This mask partitions the Fourier plane into a 100 concentric, equal-area rings. A subset of $k=100/C$ rings is then sampled.
This sampling is probabilistically weighted, biasing the sampling toward selecting lower-frequency rings (lower-frequency Fourier coefficients are shifted to the center of the image via the \texttt{fftshift} command). For each ring, given its mid-radius as $r$, its weight $w$ is given by:
\begin{equation*}
w = e^{-\frac{r}{2\nu^2}},
\end{equation*}
where $\nu = n/8$.
\item \textbf{Radial spoke.} This mask partitions the image into 100 radial spokes of equal angular width. A subset of $k=100/C$ spokes is then selected uniformly at random, without replacement. 
\end{itemize}
Fig.~\ref{app:fourier_masks} visualizes how these different masks look and their respective $\mathbf{y}$ measurements.

\section{Cryo-EM Dataset Details}
\label{app:datasets}

We provide full details on dataset sizes, preprocessing, and resolution settings used in our experiments.

\noindent\textbf{EMPIAR-10166.} Human 26S proteasome bound to Oprozomib~\cite{haselbach2017long}. We use particle images and pose metadata provided by the CESPED benchmark~\cite{sanchez-garcia2024cesped}, with 190,904 training and 23,863 validation particles. Images were originally 284 $\times$ 284 pixels and downsampled to 128 $\times$ 128 using Fourier cropping via the CryoDRGN downsample utility.

\noindent \textbf{EMPIAR-10076.} E. coli large ribosomal subunit assembly intermediates~\cite{davis2016modular}. We used the dataset provided by the CryoDRGN Zenodo repository~\cite{zhong2021cryodrgn}, with 105,519 training particles and 26,380 validation particles. The original images were 320 $\times$ 320 pixels and were downsampled to 128 $\times$ 128 using Fourier cropping for all experiments.

\noindent \textbf{EMPIAR-10648.} PKM2 protein bound to a small-molecule inhibitor~\cite{saur2020fragment}. We used particle images and pose metadata from the CESPED benchmark~\cite{sanchez-garcia2024cesped}, with 187,964 particles for training and 23,496 for validation. Images were provided at 222 $\times$ 222 pixels and were upsampled to 256 $\times$ 256 via bicubic interpolation prior to model input to match the dimensional requirements of our DDPM framework. We verified that this upsampling had no adverse effect on the structural information, as comparable 3D volume resolutions were obtained before and after upsampling.

\noindent \textbf{EMPIAR-10786.} Substance P–Neurokinin Receptor G protein complexes~\cite{harris2022selective}. We used the particle images and pose metadata provided by the CESPED benchmark~\cite{sanchez-garcia2024cesped}, with 230,927 training particles and 28,866 validation particles. Images were downsampled from 184 $\times$ 184 to 128 $\times$ 128 via Fourier cropping.

\noindent \textbf{EMPIAR-11526.} Small ribosomal subunit assembly intermediates in E. coli. We used the dataset released by Sun et al. \cite{sun2023ksga} via Zenodo, with 200,260 training particles and 25,032 validation particles. Images were provided at 256 $\times$ 256 pixels and were downsampled to 128 $\times$ 128 via Fourier cropping.

\section{Experimental Setup}
\label{appendix:exp}

\subsection{High-fidelity Reconstruction Setup}

For our reconstruction experiments, we selected $16$ random images from the validation set, which the DDPMs used in cryoSENSE did not see during training, and used that across all runs. For all runs, we compute LPIPS, SSIM, and PSNR and plot the mean and standard deviation.

\subsection{CryoDRGN Training Parameters}
\label{app:cryoDRGN}

All CryoDRGN experiments used CryoDRGN v3.4.0 on EMPIAR-10076 validation particles. Models are trained for 100 epochs with an 8-dimensional latent space, encoder and decoder sizes of 1024, and 3 residual layers. Batch size is set to 32. The \texttt{ResidLinearMLP} encoder and \texttt{FTPositionalDecoder} decoder architectures are used with standard settings. Pose and CTF metadata are provided as input. All training runs use multi-GPU acceleration with AMP mixed-precision.

\subsection{ModelAngelo Inference}
\label{app:modelangelo}

All atomic models are generated using ModelAngelo v1.0 with the \texttt{nucleotides} model bundle. Input cryo-EM density maps are reconstructed from EMPIAR-10648 validation particles for three cases: (1) original high-resolution images and cryoSENSE (2) sparse and (3) generative prior reconstructed images. ModelAngelo inference is performed using default parameters: box size of 64, stride of 16, batch size of 4, and a threshold of 0.05 for C$\alpha$ prediction. Three rounds of GNN-based model refinement are performed for all datasets, and the output models from the third round are used for all subsequent analyses.

\textbf{cryoSENSE Parameters.} We run cryoSENSE at downsampling level $K=2$ using $b=3$ masks (corresponding to $C = 1.33$). This configuration is selected by setting an SSIM threshold of 0.8 and choosing the generative prior setup that requires the fewest measurements.

\section{Additional Experiments}
\label{appendix:add_expts}

In this section, we add additional experiments that did not fit in the main text. 

\subsection{High-fidelity Reconstruction}

\begin{figure*}[t]
    \centering
    \includegraphics[width=1\linewidth]{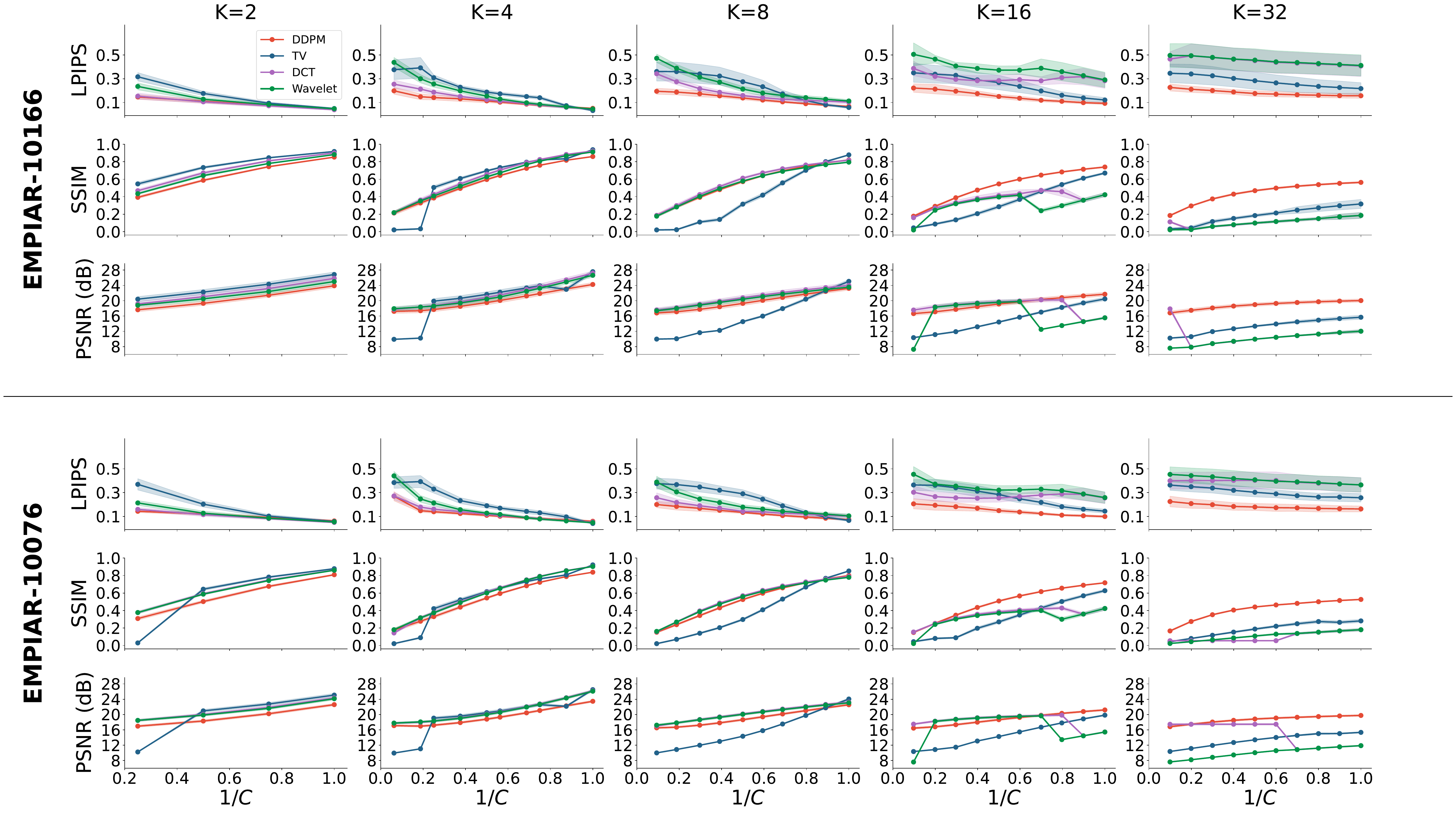}
    \caption{LPIPS, SSIM, and PSNR scores for cryoSENSE reconstructions with pixel-space masks across various levels of $K$ and $1/C$ for EMPIAR-10076 and EMPIAR-10166 (no noise $\sigma = 0.0$).}
    \label{fig:pixel_recon_1}
\end{figure*}

\begin{figure*}[t]
    \centering
    \includegraphics[width=1\linewidth]{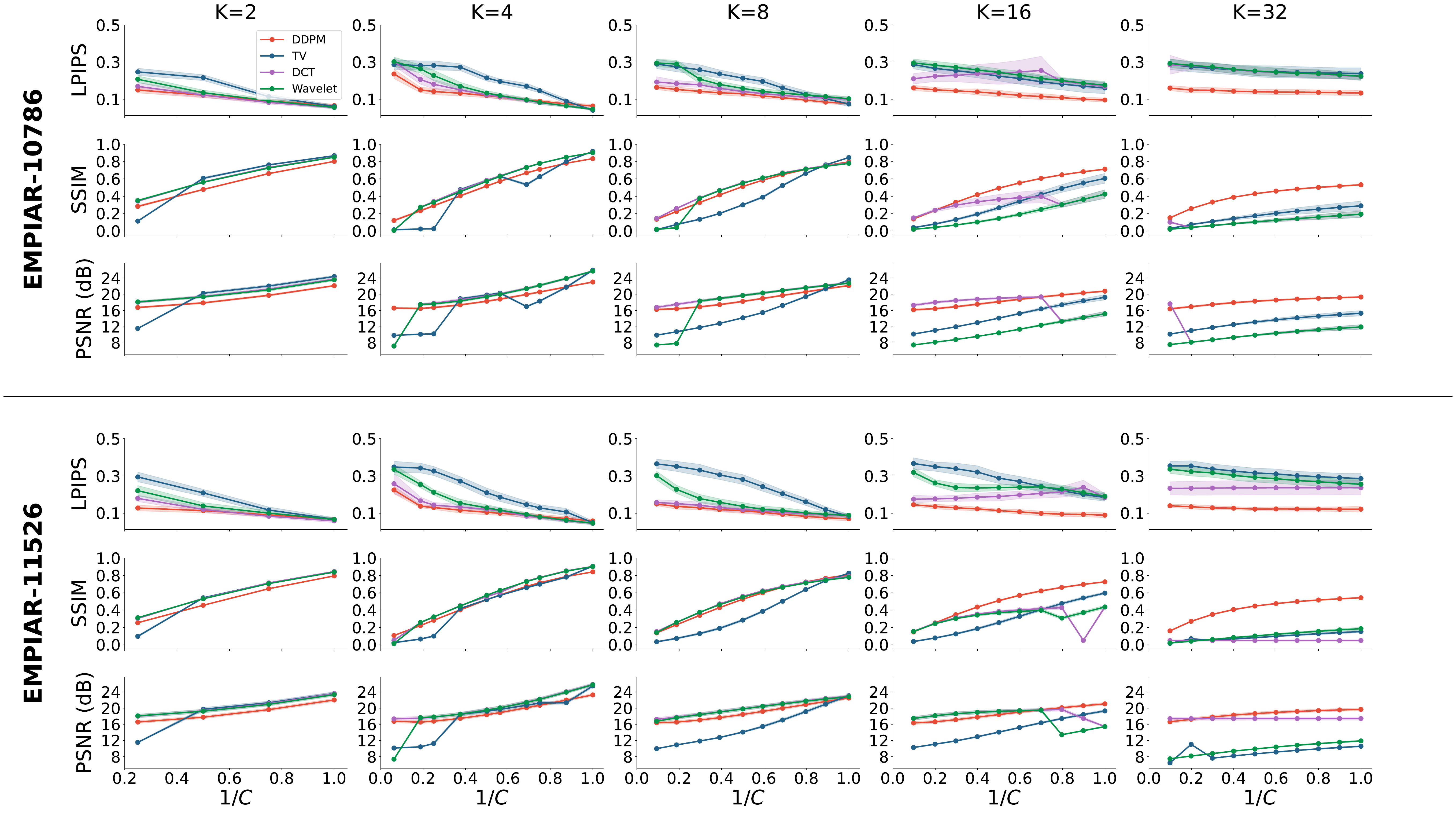}
    \caption{LPIPS, SSIM, and PSNR scores for cryoSENSE reconstructions with pixel-space masks across various levels of $K$ and $1/C$ for EMPIAR-10786 and EMPIAR-11526 (no noise $\sigma = 0.0$).}
    \label{fig:pixel_recon_2}
\end{figure*}

\begin{figure*}[t]
    \centering
    \includegraphics[width=1\linewidth]{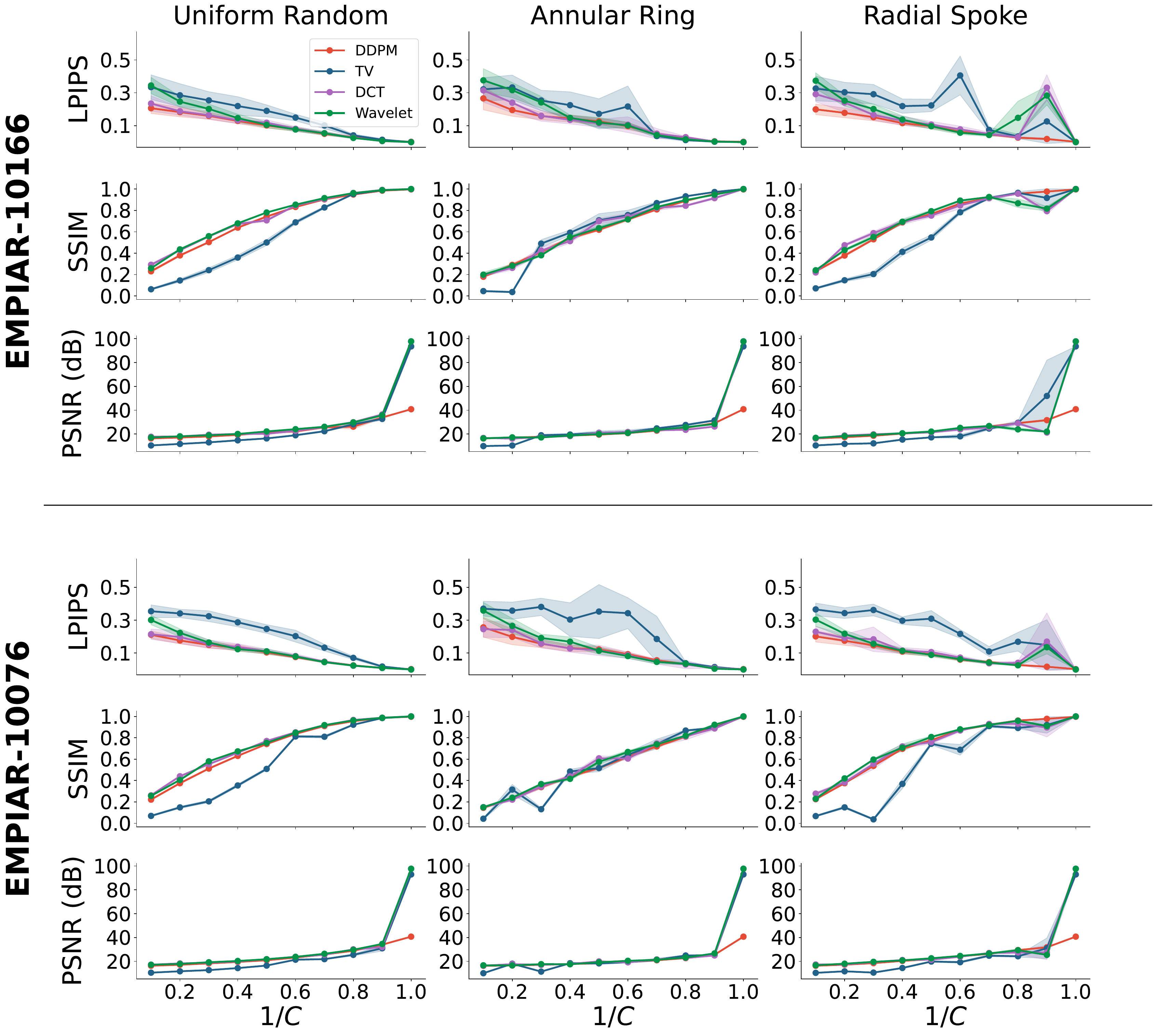}
    \caption{LPIPS, SSIM, and PSNR scores for cryoSENSE reconstructions across various Fourier masks and $1/C$ for EMPIAR-10076 and EMPIAR-10166 (no noise $\sigma = 0.0$).}
    \label{fig:fourier_recon_1}
\end{figure*}

\begin{figure*}[t]
    \centering
    \includegraphics[width=1\linewidth]{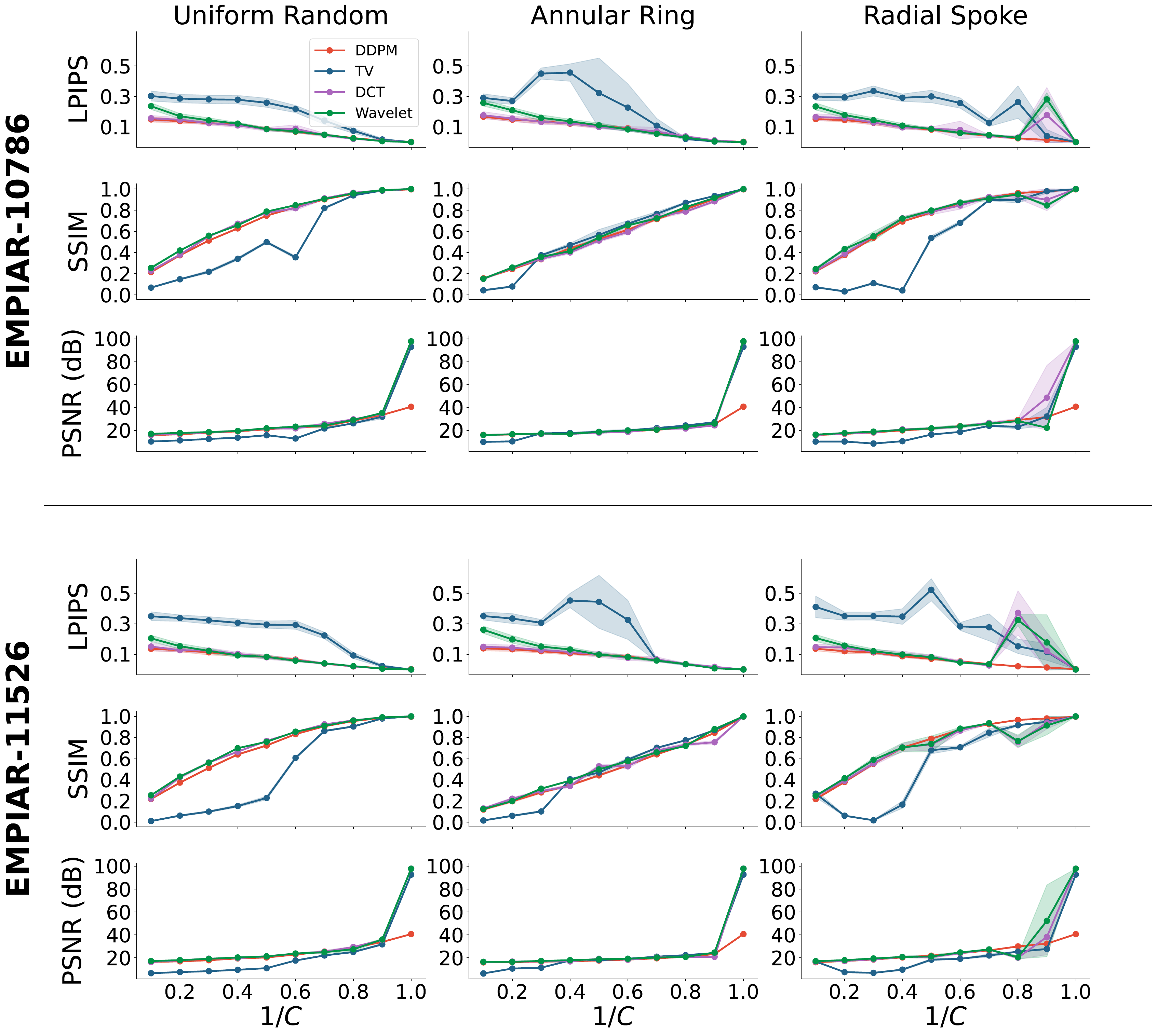}
    \caption{LPIPS, SSIM, and PSNR scores for cryoSENSE reconstructions across various Fourier masks and $1/C$ for EMPIAR-10786 and EMPIAR-11526 (no noise $\sigma = 0.0$).}
    \label{fig:fourier_recon_2}
\end{figure*}

Figs.~\ref{fig:pixel_recon_1} and~\ref{fig:pixel_recon_2} detail reconstruction performance with pixel-space masking, while Figs.~\ref{fig:fourier_recon_1} and~\ref{fig:fourier_recon_2} detail reconstruction performance with Fourier-space masking. Since the compression factor in pixel-space masking also depends on $K$, we opt to plot reconstruction performance sweeping across values of $1/C$, which ensures our plots are normalized between 0 and 1. $1/C$ can be interpreted as the ratio of measurements needed to recover the original image, where $1$ corresponds to no compression being done. For the tables in the main text, we extract average LPIPS, SSIM, and PSNR values at evenly spaced points of $1/C$.

We observe similar trends to that reported in the main text, demonstrating that cryoSENSE is able to generalize to other proteins. Across all pixel-space and Fourier-space experiments, we observe that DCT is the best performing sparse prior on average. Therefore, we use DCT for all 3D reconstruction tasks. In Fourier-space masking experiments, we additionally observe that uniform masks exhibit lower LPIPS and higher SSIM scores on average compared to annular ring and radial spoke masks. 

\subsection{High-fidelity Noisy Reconstruction}

\begin{figure*}[t]
    \centering
    \includegraphics[width=1\linewidth]{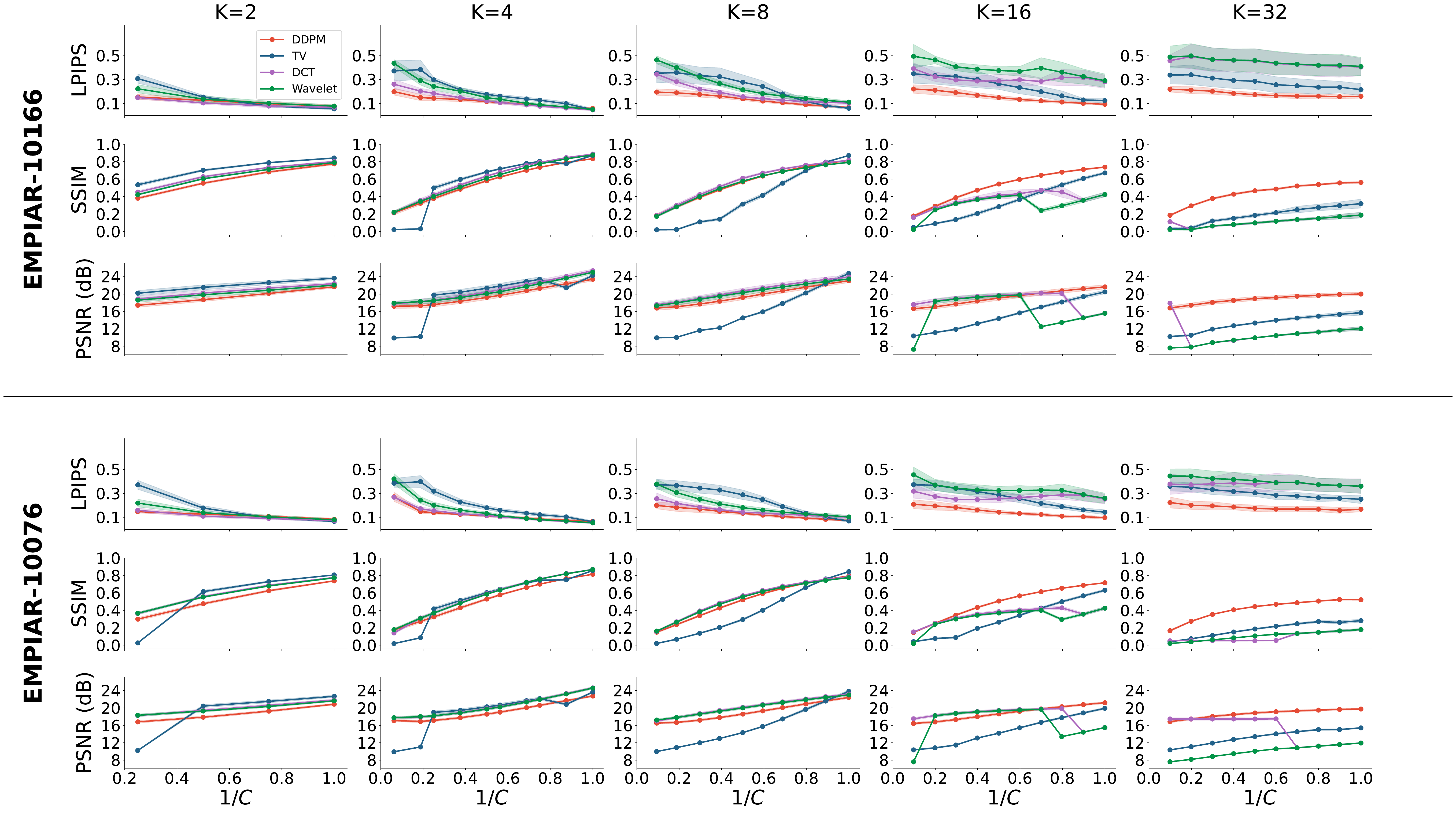}
    \caption{LPIPS, SSIM, and PSNR scores for cryoSENSE reconstructions with pixel-space masks across various levels of $K$ and $1/C$ for EMPIAR-10076 and EMPIAR-10166 in the presence of noise ($\sigma = 0.01$).}
    \label{fig:pixel_recon_1_noisy}
\end{figure*}

\begin{figure*}[t]
    \centering
    \includegraphics[width=1\linewidth]{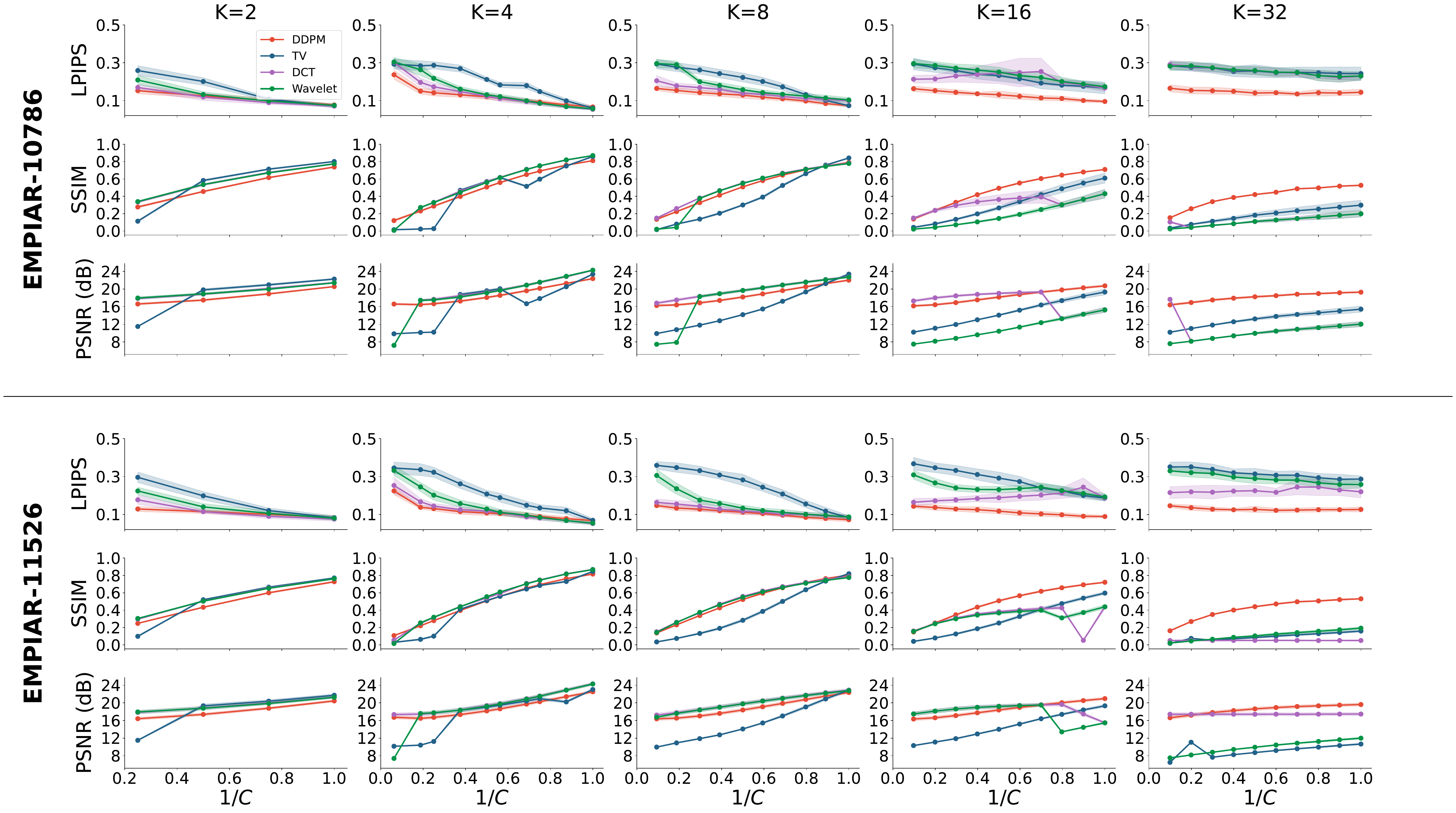}
    \caption{LPIPS, SSIM, and PSNR scores for cryoSENSE reconstructions with pixel-space masks across various levels of $K$ and $1/C$ for EMPIAR-10786 and EMPIAR-11526 in the presence of noise ($\sigma = 0.01$).}
    \label{fig:pixel_recon_2_noisy}
\end{figure*}

\begin{figure*}[t]
    \centering
    \includegraphics[width=1\linewidth]{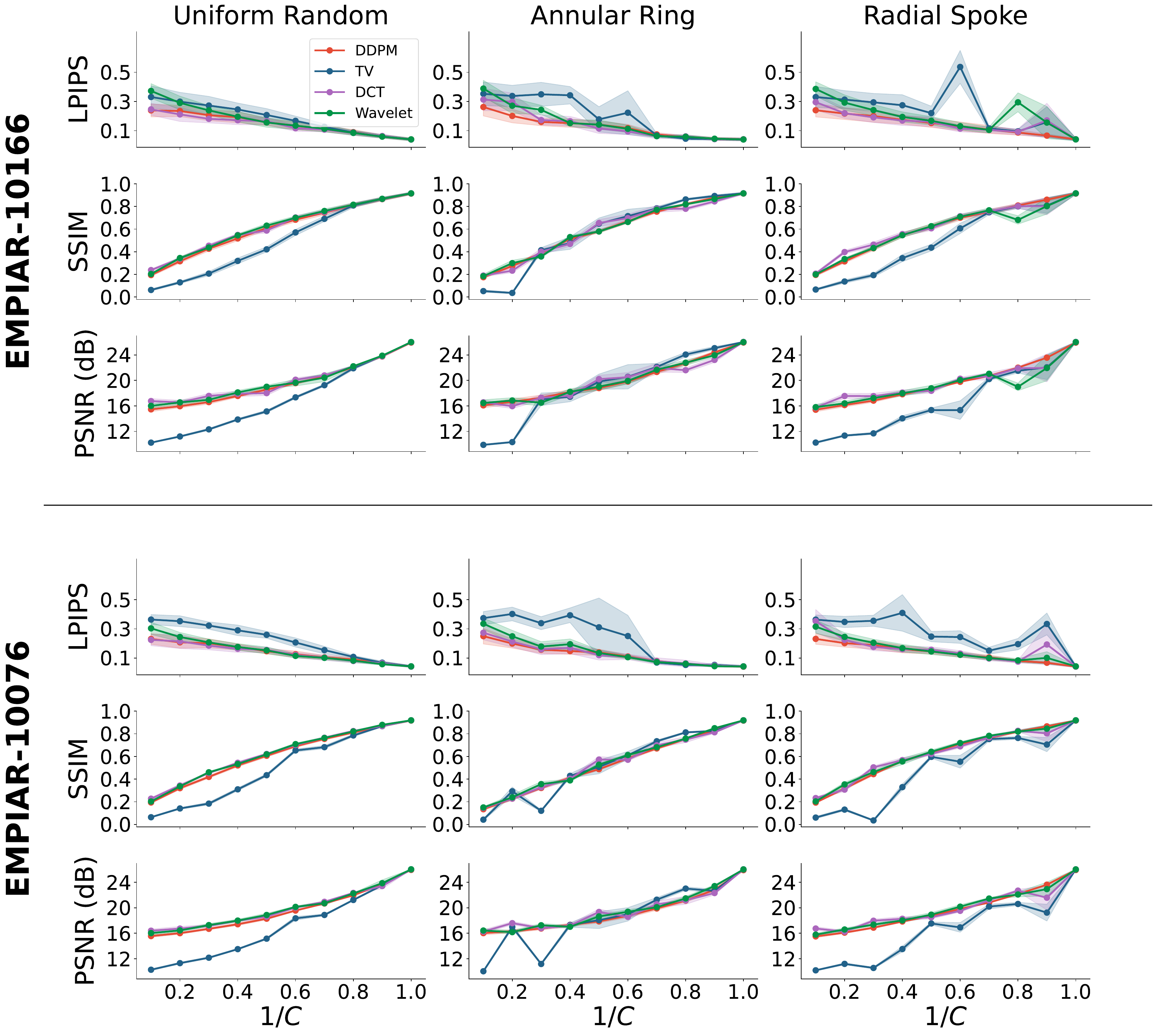}
    \caption{LPIPS, SSIM, and PSNR scores for cryoSENSE reconstructions across various Fourier masks and $1/C$ for EMPIAR-10076 and EMPIAR-10166 in the presence of noise ($\sigma = 0.01$).}
    \label{fig:fourier_recon_1_noisy}
\end{figure*}

\begin{figure*}[t]
    \centering
    \includegraphics[width=1\linewidth]{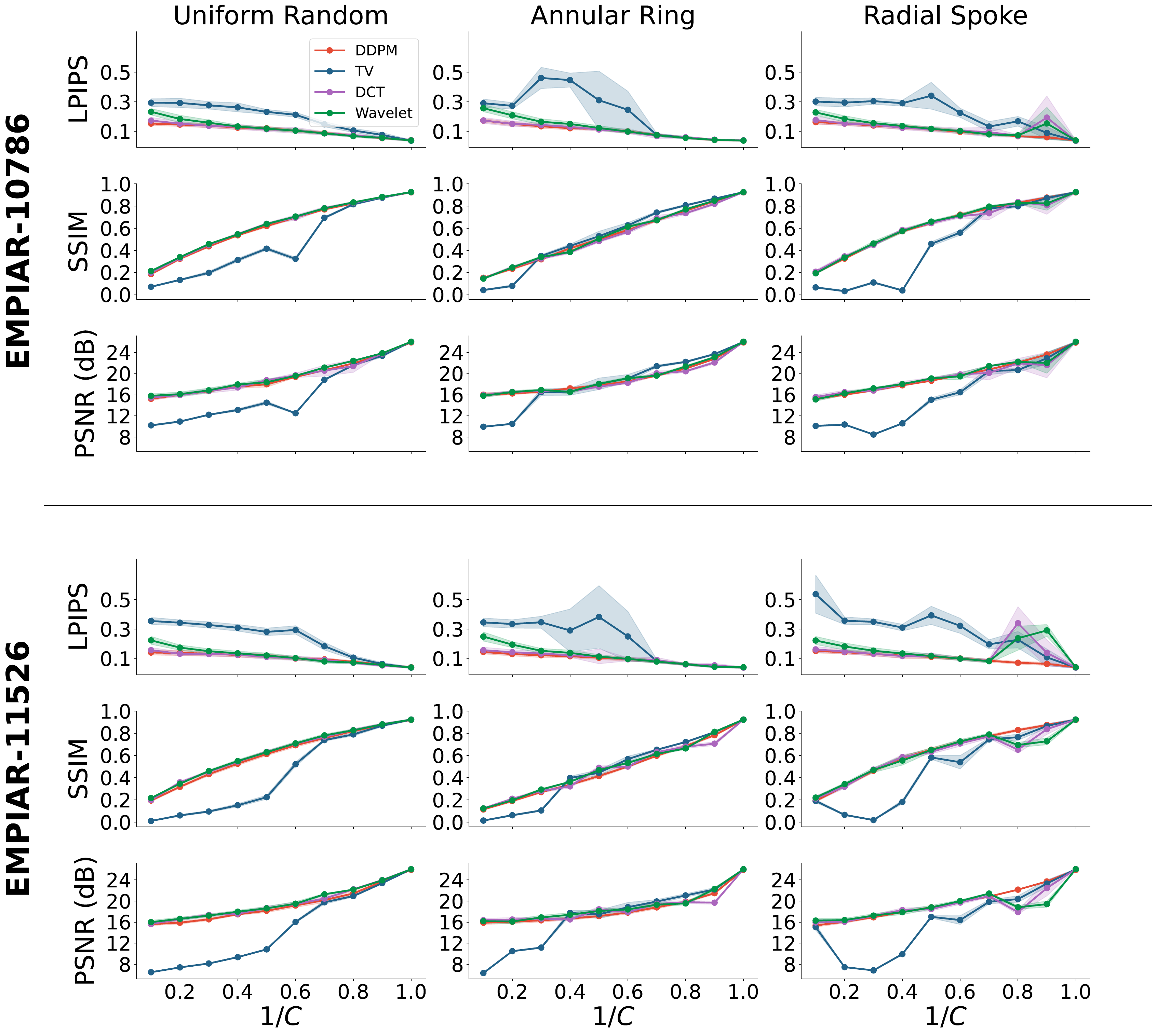}
    \caption{LPIPS, SSIM, and PSNR scores for cryoSENSE reconstructions across various Fourier masks and $1/C$ for EMPIAR-10786 and EMPIAR-11526 in the presence of noise ($\sigma = 0.01$).}
    \label{fig:fourier_recon_2_noisy}
\end{figure*}

We repeat cryoSENSE reconstructions in the noisy case, where $\sigma^2=0.01$. Figs.~\ref{fig:pixel_recon_1_noisy} and~\ref{fig:pixel_recon_2_noisy} detail reconstruction performance with pixel-space masking, while Figs.~\ref{fig:fourier_recon_1_noisy} and~\ref{fig:fourier_recon_2_noisy} detail reconstruction performance with Fourier-space masking. We observe the same trends as those made in the main text, where sparse priors are consistently better than generative priors in Fourier-space masking and under lower levels of compression, whereas generative priors are better than sparse priors in pixel-space masking and under extreme levels of downsampling and compression. This is unsurprising, as sparse and generative priors are known to be robust to moderate levels of noise~\cite{daubechies2004iterative, chung2023diffusion}.

\subsection{FSC Curves for 3D Reconstruction of EMPIAR-10076}

\begin{figure*}[t]
    \centering
    \includegraphics[width=1\linewidth]{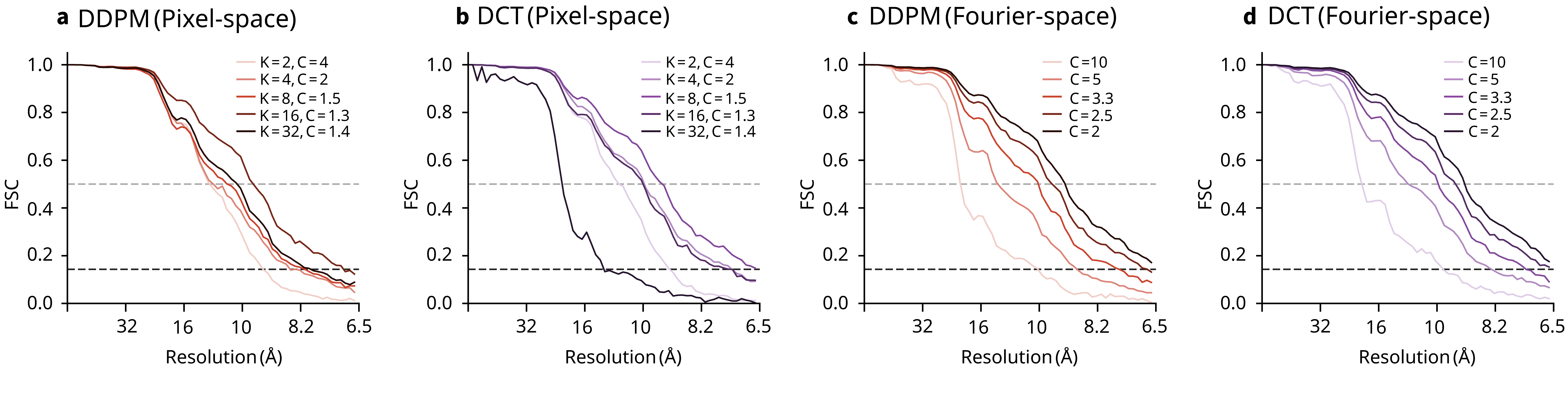}
    \caption{FSC curves for cryoSENSE reconstructions on EMPIAR-10076 with pixel-space for $\textbf{a,}$ DDPM and $\textbf{b,}$ DCT and Fourier-space masking for $\textbf{c,}$ DDPM and $\textbf{d,}$ DCT.}
    \label{fig:fsc}
\end{figure*}

Fig.~\ref{fig:fsc} presents the FSC curves quantitatively comparing DDPM and DCT reconstructions across both pixel-space and Fourier-space masks, which were summarized in Table 3 of the main text. Similar to our 2D analysis, for pixel-space masking, we note that DDPM outperforms DCT in $K=16$ and 32 downsampling regimes. For Fourier masking, DCT generally outperforms DDPM.

\textcolor{black}{\subsection{Comparison with Baseline Method}}
\label{sec:dmplug_supp}

\begin{table}[h]
\centering
\footnotesize
\setlength{\tabcolsep}{1.5pt}
\renewcommand{\arraystretch}{0.9}
\caption{Downstream validation: cryoSENSE (DDPM) vs.\ DMPlug.}
\label{tab:dmplug}
\begin{tabular}{l c c c c}
\toprule
Method & Heterogeneity Acc. $\uparrow$ & Chains $\uparrow$ & RMSD (\AA) $\downarrow$ & Confidence $\uparrow$ \\
\midrule
cryoSENSE & \textbf{82.1--91.4\%} & \textbf{42} & \textbf{2.34} & \textbf{62.5} \\
DMPlug & Random & 28 & 2.75 & 50.4 \\
\bottomrule
\end{tabular}
\end{table}

\textcolor{black}{
We compare cryoSENSE (DDPM) against DMPlug~\cite{wang2024dmplug}, a diffusion-based super-resolution (SR) method. While SR methods can produce visually smooth images, they do not enforce consistency with compressed measurements, which leads to structural inaccuracies under aggressive undersampling. We evaluate both methods on the downstream tasks presented in the main text: conformational heterogeneity recovery via CryoDRGN and atomic model building via ModelAngelo. As shown in Table~\ref{tab:dmplug}, cryoSENSE outperforms DMPlug across all metrics.}

\textcolor{black}{\subsection{Noise Model and Robustness Analysis}}
\label{sec:noise_supp}

\textcolor{black}{To evaluate noise robustness, we sweep measurement noise across SNR levels from 30 to $-20$~dB on EMPIAR-10076 at 128$\times$128 resolution, using 16 validation images. We select one representative configuration for each masking type: pixel-space masking with $K=4$, $C=2$ and Fourier-space masking with $C=2.5$ (uniform subsampling). We define the measurement SNR as $\text{SNR} = 10 \log_{10}(\text{Var}(\mathbf{y}_0) / \sigma^2)$, where $\mathbf{y}_0 = \mathcal{A}(\mathbf{x}_{\mathrm{exp}})$ denotes the compressed measurement obtained from the experimentally acquired cryo-EM image before adding Gaussian measurement noise, and $\mathrm{Var}(\mathbf{y}_0)$ is the average variance of these pre-noise measurements across the 16 images. Because the measurement operator $\mathcal{A}$ differs between pixel-space and Fourier-space masking, the variance of $\mathbf{y}_0$ differs substantially between the two configurations: $\text{Var}(\mathbf{y}_0) \approx 2.38$ for pixel-space and $\text{Var}(\mathbf{y}_0) \approx 0.027$ for Fourier-space. To ensure a fair comparison, we calibrate $\sigma$ for each case such that the same SNR in dB corresponds to the same ratio of signal power to noise power: $\sigma = \sqrt{\text{Var}(\mathbf{y}_0) / 10^{\text{SNR}/10}}$.}

\begin{figure*}[t]
    \centering
    \includegraphics[width=1\linewidth]{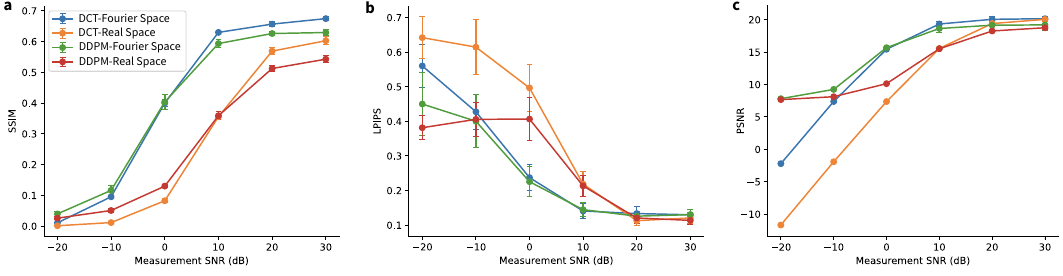}
    \caption{Noise robustness analysis on EMPIAR-10076. \textbf{a}, SSIM, \textbf{b}, LPIPS, and \textbf{c}, PSNR as a function of measurement SNR for DCT and DDPM priors under pixel-space masking ($K=4$, $b=8$, $C=2$) and Fourier-space masking ($C=2.5$, uniform subsampling). Error bars indicate standard deviation across 16 validation images. Noise levels are calibrated to the variance of the noiseless measurements for each masking type to ensure comparable SNR across configurations.}
    \label{fig:noise_sweep}
\end{figure*}

\textcolor{black}{Fig.~\ref{fig:noise_sweep} shows LPIPS, SSIM, and PSNR as a function of measurement SNR for both DCT and DDPM priors under pixel-space and Fourier-space masking. The corresponding noise standard deviations range from $\sigma = 0.049$ (30~dB) to $\sigma = 15.4$ ($-20$~dB) for pixel-space masking and from $\sigma = 0.005$ (30~dB) to $\sigma = 1.65$ ($-20$~dB) for Fourier-space masking. At high SNR ($\geq 20$~dB), all four methods converge to similar reconstruction quality. At low SNR ($\leq -10$~dB), the DDPM prior is more robust than DCT, particularly in pixel-space masking where DCT-Real degrades sharply while DDPM-Real maintains substantially lower LPIPS and higher PSNR. In Fourier-space masking, DDPM also outperforms DCT at low SNR, though both degrade. At intermediate SNR (0--10~dB), Fourier-space masking consistently outperforms pixel-space masking for both priors, and DCT pixel-space performance recovers to match DDPM pixel-space.}

\newpage
\newpage
\newpage
\newpage
\newpage

\newpage
\newpage

\newpage
\newpage

